\documentclass[reqno,10pt,a4paper]{amsart}

\usepackage{amssymb,mathptmx,psfrag,eucal,array,setspace,color,geometry,enumitem,cite}
\usepackage{graphicx}
\usepackage{tikz}

\DeclareSymbolFont{largesymbols}{OMX}{zplm}{m}{n} 

\numberwithin{equation}{section}

\newcolumntype{C}{>{$}c<{$}} 

\allowdisplaybreaks

\newcommand{\alg}[1]{\mathfrak{#1}} 
\newcommand{\grp}[1]{\mathsf{#1}} 

\newcommand{\func}[2]{#1 \left( #2 \right)} 
\newcommand{\tfunc}[2]{#1 \bigl( #2 \bigr)} 

\newcommand{\brac}[1]{\left( #1 \right)}

\newcommand{\sqbrac}[1]{\left[ #1 \right]}

\newcommand{\abs}[1]{\left\lvert #1 \right\rvert}

\newcommand{\ZZ}{\mathbb{Z}}

\newcommand{\RR}{\mathbb{R}}

\newcommand{\dd}{\mathrm{d}}   
\newcommand{\ii}{\mathfrak{i}} 
\newcommand{\ee}{\mathsf{e}}   

\newcommand{\affine}[1]{\widehat{#1}}
\newcommand{\SLG}[2]{\grp{#1} \bigl( #2 \bigr)}                             
\newcommand{\SLA}[2]{\alg{#1} \bigl( #2 \bigr)}                             
\newcommand{\AKMA}[2]{\affine{\alg{#1}} \left( #2 \right)}                  

\newcommand{\SingAlg}[1]{\mathsf{I} \bigl( #1 \bigr)}                          
\newcommand{\TripAlg}[1]{\mathsf{W} \bigl( #1 \bigr)}                          
\newcommand{\LatAlg}[1]{\mathsf{V} \bigl( #1 \bigr)}                           

\newcommand{\Fock}[1]{\mathcal{F}_{#1}}        
\newcommand{\FF}[1]{\mathcal{F}_{#1}}          
\newcommand{\ExtFock}[2]{\mathbb{F}_{#1}^{#2}} 

\newcommand{\SingKer}[1]{\mathcal{K}_{#1}}  
\newcommand{\SingIm}[1]{\mathcal{I}_{#1}}   

\newcommand{\TripIrr}[2]{\mathcal{W}_{#1}^{#2}}  

\newcommand{\VirIrr}[1]{\mathcal{L}_{#1}}   
\newcommand{\VirIrrNM}[1]{\mathcal{L}(#1)}  

\DeclareMathOperator{\tr}{tr}
\newcommand{\traceover}[1]{\tr_{\raisebox{-3pt}{$\scriptstyle #1$}}}
\newcommand{\chmap}{\mathrm{ch}}
\newcommand{\ch}[1]{\chmap \bigl[ #1 \bigr]}     

\newcommand{\modS}{\mathsf{S}} 

\newcommand{\Smat}[2]{\modS \Bigl[ #1 \ra #2 \Bigr]}      
\newcommand{\tSmat}[2]{\modS \bigl[ #1 \ra #2 \bigr]}

\newcommand{\fuse}{\mathbin{\times}}                                            
\newcommand{\extfuse}{\mathbin{\boxtimes}}                                      
\newcommand{\fuscoeff}[3]{\mathsf{N}_{#1 \, #2}^{\hphantom{#1 \, #2} \FF{#3}}}  

\newcommand{\Gr}[1]{\bigl[ #1 \bigr]}                               

\newcommand{\SingVerRing}[1]{\mathbb{V} \bigl[ \SingAlg{#1} \bigr]} 
\newcommand{\SingAtypVerRing}[1]{\mathbb{V}_{\text{atyp.}} \bigl[ \SingAlg{#1} \bigr]}
\newcommand{\TripVerRing}[1]{\mathbb{V} \bigl[ \TripAlg{#1} \bigr]}

\newcommand{\ra}{\rightarrow}
\newcommand{\lra}{\longrightarrow}
\newcommand{\dses}[5]{0 \lra #1 \overset{#2}{\lra} #3 \overset{#4}{\lra} #5 \lra 0} 

\newcommand{\eqnref}[1]{Equation~\eqref{#1}}
\newcommand{\eqnDref}[2]{Equations~\eqref{#1} and \eqref{#2}}
\newcommand{\secref}[1]{Section~\ref{#1}}

\newcommand{\appref}[1]{Appendix~\ref{#1}}

\newcommand{\cft}{conformal field theory}
\newcommand{\cfts}{conformal field theories}

\newcommand{\lcft}{logarithmic conformal field theory}
\newcommand{\lcfts}{logarithmic conformal field theories}
\newcommand{\WZW}{Wess-Zumino-Witten}

\newcommand{\hws}{highest weight vector}

\newcommand{\voa}{vertex operator algebra}
\newcommand{\voas}{vertex operator algebras}
\newcommand{\subvoa}{vertex operator subalgebra}
\newcommand{\subvoas}{vertex operator subalgebras}

\newcommand{\eps}{\varepsilon}

\renewcommand{\Im}{\operatorname{Im}}

\DeclareMathOperator{\im}{im}

\theoremstyle{plain}

\begin{document}

\title[Modular Transformations and Verlinde Formulae for Logarithmic $(p_+,p_-)$-Models]{Modular Transformations and Verlinde Formulae \\ for Logarithmic $(p_+,p_-)$-Models}

\author[D Ridout]{David Ridout}

\address[David Ridout]{
Department of Theoretical Physics \\
Research School of Physics and Engineering;
and
Mathematical Sciences Institute;
Australian National University \\
Acton, ACT 2600 \\
Australia
}

\email{david.ridout@anu.edu.au}

\author[S Wood]{Simon Wood}

\address[Simon Wood]{
Kavli Institute for the Physics and Mathematics of the Universe\\
The University of Tokyo\\
1-5, Kashiwanoha 5-Chome
Kashiwa-shi, Chiba 277-8583\\
Japan
}

\email{simon.wood@ipmu.jp}

\thanks{\today}

\begin{abstract}
The $(p_+,p_-)$ singlet algebra is a vertex operator algebra that is strongly generated by a Virasoro field of central charge $1-6\brac{p_+-p_-}^2/p_+p_-$ and a single Virasoro primary field of conformal weight $\brac{2p_+-1} \brac{2p_--1}$.  Here, the modular properties of the characters of the uncountably many simple modules of each singlet algebra are investigated and the results used as the input to a continuous analogue of the Verlinde formula to obtain the ``fusion rules'' of the singlet modules.  The effect of the failure of fusion to be exact in general is studied at the level of Verlinde products and the rules derived are lifted to the $(p_+,p_-)$ triplet algebras by regarding these algebras as simple current extensions of their singlet cousins.  The result is a relatively effortless derivation of the triplet ``fusion rules'' that agrees with those previously proposed in the literature.
\end{abstract}

\maketitle

\onehalfspacing

\section{Introduction} \label{sec:Intro}

The $(1,p)$ singlet and triplet models (for $p \geqslant 2$ a positive integer) are perhaps the most basic known examples of \lcfts{}.  Introduced in \cite{KauExt91}, their logarithmic nature was exposed, at least for $p=2$, through a connection to symplectic fermions and $bc$-ghosts \cite{GurLog93,KauCur95}.  Further investigations, for example \cite{FloMod96,Gaberdiel:1996np,Gaberdiel:1998ps,Fuchs:2003yu}, addressed the issue of generalising the tools familiar from rational \cft{} to these models and they have remained popular objects of study ever since.

Generalisations of these triplet models, called $(p_+,p_-)$-models, for $p_+$ and $p_-$ coprime and positive, were introduced in \cite{FeiLog06}.  Because their central charges match those of the Virasoro $(p_+,p_-)$ minimal models when $p_+$ and $p_-$ are greater than $1$, they are sometimes referred to as logarithmic minimal models.  One might hope that these logarithmic models could capture the universal features of critical lattice models that are missed by the minimal models (crossing probabilities, fractal dimensions and so on), but this is still contentious.  Nevertheless, there has been persistent interest in these models from both theoretical physicists and mathematicians.  One reason for this interest is that the underlying \voas{} are not simple, so these models allow one to explore the consequences of this non-simplicity in a tractable, though still very challenging, setting.  We remark that the simple quotients are precisely the minimal model \voas{}.

As with other logarithmic models, one of the main difficulties to surmount is that of obtaining a detailed structural understanding of the reducible but indecomposable modules which appear in the spectrum.  While a complete classification of the indecomposables may well be infeasible, a first aim would be to identify the spectrum of simple modules and their projective covers.  This is expected to be sufficient to construct bulk state spaces with modular invariant partition functions, for example.  However, the current state of knowledge regarding projectives in non-semisimple module categories over \voas{} is still in its infancy, so much of our intuition stems from examples like the $(p_+,p_-)$-models.

But even in examples, the rigorous identification of projectives remains a formidable task.  Indeed, this has only been achieved for $p_+ = 1$ \cite{NagTri11}.  However, the literature contains many proposals and conjectures for general $p_+$ and $p_-$ (with varying degrees of structural detail), see \cite{FeiLog06,RasWEx09,GabFus09,Wood:2009ub,AdaExp12} for example.  These proposals rely on conjectured equivalences of categories, numerical computations within integrable lattice discretisations, and explicit construction, the latter giving the most information (but requiring the most effort).  In this direction, a powerful tool for structural investigations of indecomposables is the celebrated Nahm-Gaberdiel-Kausch algorithm \cite{Gaberdiel93,NahQua94,Gaberdiel:1996kx} that explicitly constructs (filtered quotients of) the fusion product of two modules.

Determining fusion rules is, of course, another of the main difficulties one
would like to overcome along the path to understanding a given \lcft{}.  While
the Nahm-Gaberdiel-Kausch fusion algorithm allows one to construct enough of a fusion product to identify it completely in principle, the calculations are too computationally intensive for all but the smallest theories, even when performed by computer.  Another issue is that the algorithm in practice only provides an ``upper bound'' on the fusion product in the sense that the true result could be, in principle, a proper subspace of what has been deduced.  However, both of these issues can be circumvented by generalising another standard tool from rational \cft{} to the logarithmic setting:  the Verlinde formula.

The Verlinde formula \cite{VerFus88} for rational \cfts{} computes the fusion product of two modules from the modular transformation properties of their characters.  As this formula may be shown to follow from the internal consistency conditions that must be satisfied by any conformally-invariant quantum field theory \cite{MooPol88}, one expects that it should remain valid in some form for more general classes of non-rational theories.  In the logarithmic setting, characters cannot distinguish between reducible but indecomposable modules and the direct sum of their simple composition factors, hence the Verlinde formula cannot be expected to compute the true fusion rules, but only tell us which composition factors appear, and with what multiplicity, in a given fusion product.  However, this is already very valuable information.  In many cases, one can easily rule out the possibility that the simple factors combine to form an indecomposable and then the Verlinde formula gives the fusion rules as in rational theories.\footnote{We are implicitly assuming here that fusing with a fixed module defines an exact functor on our module category.}  If an indecomposable can be formed, then this formula provides the character of the indecomposable effortlessly, thus solving the ``upper bound'' problem.  Moreover, it also tells us which fusion products need to be checked for indecomposability, thus potentially saving computational resources.

Unfortunately, the modular properties of the triplet $(p_+,p_-)$-models (with
$(p_+,p_-) \neq (1,1)$) are not as nice as one could have hoped for.  In
particular, the character of the vacuum module transforms under $\modS \colon
\tau \to -1/\tau$ into a linear combination of other characters, but the
coefficients depend non-trivially on $\tau$ \cite{FloMod96,FeiLog06}.  This
would appear to invalidate a na\"{\i}ve application of the Verlinde formula.
Nevertheless, one can arrive at a $\tau$-independent $\modS$-transformation by
postulating a non-standard automorphy matrix and a generalised Verlinde
formula exploiting this has been demonstrated for the triplet models with $p_+
= 1$ \cite{Fuchs:2003yu}.  This proposed recipe does produce non-negative
integer structure constants which agree with the known (Grothendieck) fusion
coefficients.  However, the generalised Verlinde formula itself is
significantly more unwieldy than the original and we are not aware of any
attempts to derive its analogues for other \lcfts{}. Another way of obtaining
\(\tau\)-independent coefficients is to enlarge the space of characters to
the space of torus 1-point functions, that is, to add certain linear
combinations of characters multiplied by appropriate powers of \(\tau\)
\cite{MiyMod04}, see also \cite{FloMod96,FloLog06,FeiLog06}.  Other proposals 
for triplet Verlinde formulae may be found in \cite{FloVer07,Gaberdiel:2007jv,
GaiRad09,PeaGro10,RasFus10}.

Here, we follow a different path to the Verlinde formula.  Instead of working
directly with the triplet $(p_+,p_-)$-models that have received so much attention in the literature, we focus our attention on the relatively unexplored
singlet $(p_+,p_-)$-models.  Whereas the triplet \voas{} are known to possess
a finite number of inequivalent simple modules
\cite{EhoHow93,FloMod96,FeiLog06,AdaTri08,AdaMil10,TsuExt13}, the singlet
algebras admit an uncountable infinity of
them.
However, this is not a bug, but a feature!  We will see that the modular transformation properties of the characters of these simple singlet modules are very well behaved.  Moreover, applying the standard Verlinde formula (but with an integral replacing the sum) leads again to non-negative integer structure constants.  Finally, these results can be lifted from the singlet algebra to its triplet cousin using the technology of simple current extensions.  In particular, our results provide an effortless derivation of the $(1,p)$ triplet (Grothendieck) fusion rules without the need for non-standard automorphy factors and complicated generalisations of the Verlinde formula.

This path to the Verlinde formula is actually a special case of a rather more
general formalism that has been proposed for non-rational \cfts{} in
\cite{CreLog13}.  There, one starts with a continuous spectrum of so-called
\emph{standard} modules which are typically simple and whose characters have
good modular properties.  In logarithmic theories, the \emph{atypical}
standard modules are reducible but indecomposable and the not-so-good modular
properties of the characters of the simple subquotients may be determined
using standard methods of homological algebra.  This approach was developed
for logarithmic models based on affine (super)algebras
\cite{RozQua92,QueFre07,CreRel11,CreMod12} where the natural spectrum is
continuous. One of the successes of this approach is the complete resolution
of the longstanding problem of negative fusion coefficients in fractional level \WZW{} models \cite{RidSL208,CreMod13}.

The application of this general formalism to the $(1,p)$ singlet models is relatively straightforward (see \cite[Sec.~3]{CreLog13} for the case $p = 2$ and \cite{CreFal13} for more general $p$).  However, the generalisation to all $(p_+,p_-)$ singlet algebras is rather more interesting because, in the case where $p_+$ and $p_-$ are both greater than $1$, the fusion product is no longer expected to define exact functors on the (natural) category of \voa{} modules.  This non-exactness was first noted in \cite{GabFus09} for the $(2,3)$ triplet model.  Consequently, the Grothendieck group spanned by the (equivalence classes of) simple modules does not inherit a ring structure from the fusion product.  One therefore cannot expect that the ring structure defined by the Verlinde formula on the span of the characters of the simple modules --- we call the resulting object the \emph{Verlinde ring} --- will coincide with a Grothendieck ring of fusion.  The natural question of how the non-exactness of fusion is manifested in the Verlinde ring is what motivated our work on this problem.\footnote{A second motivation is to study the modular story for examples of logarithmic theories (see also \cite{BabTak12} in this regard) involving indecomposables that are structurally more complicated than those of the $(1,p)$-models.}  As we will see, the answer is natural and satisfying, though there are subtleties worth remarking upon.

\medskip

We begin in \secref{sec:Background} with notations and conventions, reviewing
the definitions of the singlet and triplet $(p_+,p_-)$-algebras in terms of
the Heisenberg algebra and its simple current extensions.  The irreducible
modules of both \voas{} are constructed and their classifications are quoted
with the necessary structural aspects of these modules being deferred to an
appendix.  This material has many sources, for example \cite{FeiRep90,IohRep11}. Here, we
mostly follow the notation of \cite{TsuExt13}.

In \secref{sec:SingFus}, the modular properties of the characters of the
singlet modules are derived.  We begin with the standard modules which are of
Feigin-Fuchs type, adding a Heisenberg charge to their characters so that all non-isomorphic simple modules have distinct characters.  The $\modS$-transformations of these characters are deduced in the usual way.  The algebraic definitions of the remaining (atypical) simple modules then lead to resolutions for each atypical simple in terms of standards.  The resulting character formulae then give the $\modS$-transformations of the atypical characters directly.  Of note here is that when $p_+$ and $p_-$ are both greater than $1$, there exist atypical simple modules $\VirIrr{r,s}$ whose $\modS$-matrix entries cannot be expressed as functions, but only as \emph{distributions}.  Indeed, this is also the case for the (non-simple) vacuum module.

We then turn to the Verlinde formula and the Verlinde product that it induces
in \secref{sec:Verlinde}.  Most importantly, we show that the $\VirIrr{r,s}$
completely decouple in the Verlinde ring and may be consistently set to zero.
This lets us replace, when $p_+,p_- > 1$, the $\modS$-matrix entries involving
the vacuum module by those involving its maximal submodule, which happens to
be simple.  The Verlinde formula is then well-defined, because we no longer
need to divide by a distribution, and direct computation ensues. We thereby obtain a completely explicit description in \eqnref{eq:SingVerRing} of the Verlinde product of the characters of any two simple singlet modules, excepting the $\VirIrr{r,s}$ whose characters have been set to $0$.

This result is then lifted to the triplet models through their realisations as simple current extensions of the corresponding singlet models.  Actually, these realisations remain conjectural in general because we can only verify the simple current property at the level of the Verlinde rings, not the fusion rings themselves.  Nevertheless, we apply standard simple current technology to deduce the triplet analogues of \eqnref{eq:SingVerRing}.  The resulting Verlinde product rules, reported in \eqnref{eq:TripVerRingRules}, are then compared favourably with the rules that have been proposed elsewhere in the literature.  We close with a conclusion and discussion of our results.

\section{The $(p_+,p_-)$ Singlet and Triplet Models} \label{sec:Background}

In this section, we introduce and fix our notation for the singlet and triplet models.  These \cfts{} are parametrised by two coprime positive integers $p_+$ and $p_-$.  As one would expect, many of the important quantities that we will study take a somewhat unwieldy form when expressed in terms of these parameters, so to partially alleviate this, we introduce the following quantities:
\begin{subequations}
\begin{gather}
\alpha_+ = \sqrt{\frac{2p_-}{p_+}}, \qquad 
\alpha_- = -\sqrt{\frac{2p_+}{p_-}}, \qquad 
\alpha_0 = \alpha_+ + \alpha_-, \\
\alpha = p_+ \alpha_+ = -p_- \alpha_- = \sqrt{2 p_+ p_-}, \qquad 
\alpha_{r,s} = \frac{1-r}{2} \alpha_+ + \frac{1-s}{2} \alpha_-, \qquad 
\alpha_{r,s;n} = \alpha_{r,s} + \frac{1}{2} n \alpha.
\end{gather}
\end{subequations}
Here, $r$, $s$ and $n$ will always be assumed to be integers.  Note that the $\alpha_{r,s;n}$ so-defined satisfy
\begin{equation}
\alpha_{r \mp p_+,s;n} = \alpha_{r,s;n \pm 1} = \alpha_{r,s \pm p_-;n}.
\end{equation}
We may therefore choose $n$ so that $1 \leqslant r \leqslant p_+$ and $1 \leqslant s \leqslant p_-$, when convenient.

\subsection{Feigin-Fuchs Modules} \label{sec:FF}

Consider the Fock module $\Fock{\lambda}$ of the Heisenberg algebra $\AKMA{gl}{1}$ with highest weight $\lambda \in \RR$.  As is well known, the vacuum module $\Fock{0}$ carries the structure of a \voa{}.  There exists a continuous family of conformal structures for this \voa{} and we will choose the corresponding Virasoro algebra so that the central charge is
\begin{subequations}
\begin{equation}
c = 1 - 3 \alpha_0^2
\end{equation}
and the \hws{} of $\Fock{\lambda}$ has conformal weight
\begin{equation} \label{eq:FockCW}
\Delta_{\lambda} = \frac{1}{2} \lambda \brac{\lambda - \alpha_0} = \frac{1}{2} \brac{\lambda - \frac{\alpha_0}{2}}^2 - \frac{\alpha_0^2}{8}.
\end{equation}
\end{subequations}
Restricting to the action of this Virasoro algebra, 
the Fock modules $\Fock{\lambda}$ become Virasoro modules which we shall also denote by $\FF{\lambda}$.  When considering the $\FF{\lambda}$ as Virasoro modules, we shall refer to them as Feigin-Fuchs modules.

The structure of these Feigin-Fuchs modules was determined by Feigin and Fuchs in \cite{FeiRep90} (see also \cite{IohRep11} for a comprehensive treatment).  If $\lambda$ is not of the form $\alpha_{r,s;n}$ for some $r,s,n \in \ZZ$, then $\FF{\lambda}$ is simple as a Virasoro module.  If we choose $n$ in what follows so that $1 \leqslant r \leqslant p_+$ and $1 \leqslant s \leqslant p_-$, then the structure depends only upon whether $r$ and $s$ are $p_+$ and $p_-$, respectively, and upon the sign of $n$.  We defer the explicit structural details of these Virasoro modules, in the form of socle filtrations, to \appref{app:Structures}.

The structure of the Feigin-Fuchs modules may be used to derive the Felder complexes \cite{FelBRST89}
\begin{subequations} \label{eq:Felder}
\begin{gather}
\cdots \lra \FF{r,s;-2} \lra \FF{p_+ - r,s;-1} \lra \FF{r,s;0} \lra \FF{p_+ - r,s;+1} \lra \FF{r,s;+2} \lra \cdots \qquad \text{(\(r \neq p_+\)),} \label{eq:Felder+} \\
\cdots \lra \FF{r,s;+2} \lra \FF{r,p_- - s;+1} \lra \FF{r,s;0} \lra \FF{r,p_- - s;-1} \lra \FF{r,s;-2} \lra \cdots \qquad \text{(\(s \neq p_-\)),} \label{eq:Felder-}
\end{gather}
\end{subequations}
where we have simplified our notation by setting $\FF{r,s;n} \equiv
\FF{\alpha_{r,s;n}}$.  Indeed, the Virasoro homomorphisms defining these
complexes may be identified with (suitably regularised) powers of screening
operators \cite{TsuKan:1986}.
The complex \eqref{eq:Felder+} may be checked to be exact when $s=p_-$.  Moreover, it only fails to be exact when $s \neq p_-$ at the $n=0$ term, in which case the homology is the simple Virasoro module $\VirIrr{r,s}$ whose \hws{} has conformal weight $\Delta_{r,s} \equiv \Delta_{\alpha_{r,s}}$.  Similarly, \eqref{eq:Felder-} is exact for $r=p_+$ and otherwise only has non-zero homology, again given by $\VirIrr{r,s}$, at $n=0$.

\subsection{The Singlet Algebra and its Modules} \label{sec:Sing}

We define the following (Virasoro) submodules of $\FF{r,s;n}$ for $1 \leqslant r \leqslant p_+$ and $1 \leqslant s \leqslant p_-$:
\begin{equation} \label{eq:DefKI}
\begin{gathered}
\begin{aligned}
\SingKer{r,s;n}^+ &= \ker \sqbrac{\FF{r,s;n} \lra \FF{p_+ - r,s;n+1}}, \\
\SingKer{r,s;n}^- &= \ker \sqbrac{\FF{r,s;n} \lra \FF{r,p_- - s;n-1}},
\end{aligned}
\qquad
\begin{aligned}
\SingIm{r,s;n}^+ &= \im \sqbrac{\FF{p_+ - r,s;n-1} \lra \FF{r,s;n}} \\
\SingIm{r,s;n}^- &= \im \sqbrac{\FF{r,p_- - s;n+1} \lra \FF{r,s;n}}
\end{aligned}
\qquad
\begin{aligned}
&\text{(\(r \neq p_+\)),} \\
& \text{(\(s \neq p_-\)),}
\end{aligned}
\\
\SingKer{r,s;n} = \SingKer{r,s;n}^+ \cap \SingKer{r,s;n}^-, \qquad 
\SingIm{r,s;n} = \SingIm{r,s;n}^+ \cap \SingIm{r,s;n}^- \qquad \text{(\(r \neq p_+\), \(s \neq p_-\)).}
\end{gathered}
\end{equation}
If $p_+ = 1$, the $\SingKer{r,s;n}^+$ are not defined by \eqref{eq:DefKI}.  We will therefore set $\SingKer{r,s;n}^+ = \FF{r,s;n}$ in this case.  Similarly, if $p_- = 1$, we set $\SingKer{r,s;n}^- = \FF{r,s;n}$.  Because of the exactness of the Felder complexes, we have the identifications ($r \neq p_+$, $s \neq p_-$)
\begin{equation} \label{eq:Ker=Im}
\SingKer{r,p_-;n}^+ \cong \SingIm{r,p_-;n}^+, \quad 
\SingKer{p_+,s;n}^- \cong \SingIm{p_+,s;n}^- \quad \text{(for all \(n\));} \qquad
\SingKer{r,s;n}^{\bullet} \cong \SingIm{r,s;n}^{\bullet} \quad \text{(for all \(n \neq 0\)),}
\end{equation}
where the superscript ``$\bullet$'' stands for $+$, $-$, or is empty.  Working out the Virasoro module structures of the $n=0$ modules using the socle series of the Feigin-Fuchs modules (\appref{app:Structures}), one arrives at the (non-split) short exact sequences
\begin{equation}\label{es:ImKerVir}
\dses{\SingIm{r,s;0}^{\bullet}}{}{\SingKer{r,s;0}^{\bullet}}{}{\VirIrr{r,s}} \qquad \text{(\(r \neq p_+\), \(s \neq p_-\)).}
\end{equation}
Finally, the definitions \eqref{eq:DefKI} immediately imply the exact sequences
\begin{equation} \label{es:KerFFIm}
\begin{gathered}
\dses{\SingKer{r,s;n}^+}{}{\FF{r,s;n}}{}{\SingIm{p_+ - r,s;n+1}^+} \qquad \text{(\(r \neq p_+\)),} \\
\dses{\SingKer{r,s;n}^-}{}{\FF{r,s;n}}{}{\SingIm{r,p_- - s;n-1}^-} \qquad \text{(\(s \neq p_-\)),}
\end{gathered}
\end{equation}
for all $n$, and further contemplation of Virasoro structures (see \appref{app:Structures}) leads to
\begin{equation}\label{es:ImImIm}
\begin{gathered}
\dses{\SingIm{r,s;n}}{}{\SingIm{r,s;n}^+}{}{\SingIm{r,p_- - s;n-1}} \\
\dses{\SingIm{r,s;n}}{}{\SingIm{r,s;n}^-}{}{\SingIm{p_+ - r,s;n+1}}
\end{gathered}
\qquad \text{(\(r \neq p_+\), \(s \neq p_-\)),}
\end{equation}
again for all $n$, which are likewise exact.

Recall that $\FF{1,1;0} = \FF{0}$ carries the structure of a \voa{}.  As
$\SingKer{1,1;0}^+$ and $\SingKer{1,1;0}^-$ are both kernels of screening
operators acting on this \voa{}, they define \subvoas{}, as does their
intersection $\SingKer{1,1;0}$.  The \voa{} corresponding to $\SingKer{1,1;0}$
is called the \emph{singlet algebra} $\SingAlg{p_+,p_-}$.  It is simple if and
only if $p_+$ or $p_-$ is 1.  We remark that if $p_+ = p_- = 1$, then
$\SingKer{1,1;0}^+ = \SingKer{1,1;0}^- = \SingKer{1,1;0} = \FF{1,1;0}$ and we
see that the singlet algebra $\SingAlg{1,1}$ is nothing but the Heisenberg
algebra (with central charge $1$).  In general, the singlet algebra is
strongly generated by the energy-momentum tensor and a single Virasoro primary
of dimension $\brac{2p_+ - 1} \brac{2p_- - 1}$ \cite{FeiLog06,AdaMil10,TsuExt13}.

Each of the $\FF{\lambda}$, $\SingKer{r,s;n}^{\bullet}$ and $\SingIm{r,s;n}^{\bullet}$, as well as the $\VirIrr{r,s}$, become modules for the singlet \voa{}.  A complete list of simple $\SingAlg{p_+,p_-}$-modules is given by
\begin{itemize}
\item the $\FF{\lambda}$ with $\lambda \neq \alpha_{r,s;n}$ for any $r,s,n \in \ZZ$,
\item the $\FF{p_+,p_-;n}$ for all $n \in \ZZ$,
\item the $\SingIm{r,p_-;n}^+$ for all $1 \leqslant r \leqslant p_+ - 1$ and $n \in \ZZ$,
\item the $\SingIm{p_+,s;n}^-$ for all $1 \leqslant s \leqslant p_- - 1$ and $n \in \ZZ$,
\item the $\SingIm{r,s;n}$ for all $1 \leqslant r \leqslant p_+ - 1$, $1 \leqslant s \leqslant p_- - 1$ and $n \in \ZZ$,
\item the $\VirIrr{r,s} \cong \VirIrr{p_+ - r,p_- - s}$ for all $1 \leqslant r \leqslant p_+ - 1$ and $1 \leqslant s \leqslant p_- - 1$.
\end{itemize}
The case when $p_+$ or $p_-$ is $1$ was settled in
\cite{AdaCla03,AdaMil09}. When $p_+,p_-\geq 2$, then the completeness of the
above list is a straightforward corollary of \cite[Thm.~D]{TsuExt13} using the
same arguments as in \cite{AdaMil09}.
We remark that when $p_+ = 1$, the sets of $\SingIm{r,p_-;n}^+$, $\SingIm{r,s;n}$ and $\VirIrr{r,s}$ are empty --- the only $\SingAlg{1,p_-}$-simples are the $\FF{\lambda}$, the $\FF{1,p_-;n}$ and the $\SingIm{1,s;n}^-$.  The story when $p_- = 1$ is similar.

Notice that the $\FF{\lambda}$ are simple for generic $\lambda$.  In the formalism proposed in \cite{CreLog13} for general (logarithmic) \cfts{}, the $\FF{\lambda}$ may be identified as the \emph{standard} singlet modules.  The simple standard modules, those with $\lambda \neq \alpha_{r,s;n}$ or with $\lambda = \alpha_{p_+,p_-;n}$, are called \emph{typical} in this formalism and the remaining simple modules, the $\SingIm{r,s;n}^{\bullet}$ and $\VirIrr{r,s}$, are examples of \emph{atypical} singlet modules.  We will use this terminology freely in what follows.

\subsection{The Triplet Algebra and its Modules} \label{sec:Trip}

Just like the singlet algebra \(\SingAlg{p_+,p_-}\), the \emph{triplet algebra}
\(\TripAlg{p_+,p_-}\) can be defined as a \subvoa{} of a lattice \voa{} \(\LatAlg{p_+,p_-}\). 
This lattice algebra may be characterised as the simple current extension of \(\FF{1,1;0}\) 
by the group of simple currents generated by
\(\FF{1,1;2}\) (or alternatively by \(\FF{1,1;-2}\)). In terms of
Feigin-Fuchs modules, we therefore have the decomposition
\begin{align}
  \LatAlg{p_+,p_-}=\bigoplus_{k\in\mathbb{Z}} \FF{1,1;2k}.
\end{align}
This lattice \voa{} is known to be rational, meaning that all of its modules are
semisimple and that there are only finitely many inequivalent simple modules.
The number of inequivalent simple modules is, in this case,
exactly \(2p_+p_-\) and they can be parametrised by two integers, \(1\leqslant r\leqslant p_+\) and \(1\leqslant s\leqslant p_-\), and a label ``$\pm$''.  We denote these simple \(\LatAlg{p_+,p_-}\)-modules by $\ExtFock{r,s}{\pm}$.  They may be decomposed into
Feigin-Fuchs modules as follows:
\begin{align}
  \ExtFock{r,s}{+}=\bigoplus_{k\in\mathbb{Z}}\FF{r,s;2k}\,,\qquad
  \ExtFock{r,s}{-}=\bigoplus_{k\in\mathbb{Z}}\FF{r,s;2k+1}\,.
\end{align}

Just like the Feigin-Fuchs modules, the lattice modules $\ExtFock{r,s}{\pm}$ can also be arranged into
Felder complexes connected by (appropriately regularised) powers of screening operators:
\begin{subequations}
\begin{gather}
    \cdots \lra \ExtFock{r,s}{\eps} \lra \ExtFock{p_+ - r,s}{-\eps} \lra
    \ExtFock{r,s}{\eps} \lra \ExtFock{p_+ - r,s}{-\eps}
    \lra \ExtFock{r,s}{\eps} \lra \cdots \qquad \text{(\(r \neq p_+\)),}  \\
    \cdots \lra \ExtFock{r,s}{\eps} \lra \ExtFock{r,p_- - s}{-\eps} \lra \ExtFock{r,s}{\eps} \lra 
    \ExtFock{r,p_- - s}{-\eps} \lra \ExtFock{r,s}{\eps} \lra \cdots \qquad \text{(\(s \neq p_-\)).} 
\end{gather}
\end{subequations}
Here, $\eps$ stands for either ``$+$'' or ``$-$''.  As with the singlet algebra, the triplet algebra \(\TripAlg{p_+,p_-}\) may also be defined as an intersection of kernels:
\begin{align}
  \TripAlg{p_+,p_-}=\ker \sqbrac{\ExtFock{1,1}{+} \lra \ExtFock{p_+ - 1,1}{-}}
  \cap \ker \sqbrac{\ExtFock{1,1}{+} \lra \ExtFock{1,p_--1}{-}}\,.
\end{align}
Again, when $p_+ = 1$ or $p_- = 1$, at least one of the Felder complexes is
not defined and its corresponding kernel is replaced by the lattice module $\ExtFock{1,1}{+}$.  In particular, it follows that $\TripAlg{1,1} = \ExtFock{1,1}{+}$ which is well known to be isomorphic to the level $1$ \voa{} $\AKMA{sl}{2}_1$.  In general, the triplet \voa{} $\TripAlg{p_+,p_-}$ is strongly generated by the energy momentum tensor and three Virasoro primaries of dimension $\brac{2p_+ - 1} \brac{2p_- - 1}$ \cite{FeiLog06,AdaMil10,TsuExt13}.  The singlet algebra $\SingAlg{p_+,p_-}$ is naturally generated as a \subvoa{} by removing two of these Virasoro primaries.

Unlike the lattice algebra $\LatAlg{p_+,p_-}$, the triplet algebra \(\TripAlg{p_+,p_-}\) is not rational in general,\footnote{The single exception is $\TripAlg{1,1} \cong \AKMA{sl}{2}_1$ which is well known to be rational and even unitary.} because there
exist non-semisimple triplet modules \cite{Gaberdiel:1996np,AdaTri08}. However, the number of inequivalent simple
\(\TripAlg{p_+,p_-}\)-modules is finite and, in fact, this number is 
\(\tfrac{1}{2}(p_+-1)(p_--1)+2p_+p_-\) \cite{FeiLog06,AdaMil10,TsuExt13}.
We give a complete list of these simples along with their decompositions
 into singlet modules (for later convenience):
\begin{itemize}[itemsep=2mm]
\item $\TripIrr{p_+,p_-}{+}=\displaystyle\bigoplus_{k\in\ZZ} \FF{p_+,p_-;2k}\ $ and 
      $\ \TripIrr{p_+,p_-}{-}=\displaystyle\bigoplus_{k\in\ZZ} \FF{p_+,p_-;2k+1}$,
\item $\TripIrr{r,p_-}{+}=\displaystyle\bigoplus_{k\in\ZZ}
  \SingIm{r,p_-;2k}^+\ $ and 
      $\ \TripIrr{r,p_-}{-}=\displaystyle\bigoplus_{k\in\ZZ}
      \SingIm{r,p_-;2k+1}^+\ $ for all \(1\leqslant r\leqslant p_+-1\),
\item $\TripIrr{p_+,s}{+}=\displaystyle\bigoplus_{k\in\ZZ}
  \SingIm{p_+,s;2k}^-\ $ and 
      $\ \TripIrr{p_+,s}{-}=\displaystyle\bigoplus_{k\in\ZZ}
      \SingIm{p_+,s;2k+1}^-\ $ for all \(1\leqslant s\leqslant p_--1\),
\item $\TripIrr{r,s}{+}=\displaystyle\bigoplus_{k\in\ZZ} \SingIm{r,s;2k}\ $ and 
      $\ \TripIrr{r,s}{-}=\displaystyle\bigoplus_{k\in\ZZ} \SingIm{r,s;2k+1}\ $ for all \(1\leqslant r\leqslant p_+-1\) and \(1\leqslant s\leqslant p_--1\),
\item \(\VirIrr{r,s}\cong \VirIrr{p_+-r,p_--s}\ \) for all \(1\leqslant r\leqslant p_+-1\) and \(1\leqslant s\leqslant p_--1\).
\end{itemize}
Again, this is an easy consequence of the results of \cite{FeiLog06,AdaMil10,TsuExt13}.  Note that when $p_+ = 1$, there are no $\TripIrr{r,p_-}{\pm}$, $\TripIrr{r,s}{\pm}$ or $\VirIrr{r,s}$ in this list --- the simples are exhausted by the $\TripIrr{1,s}{\pm}$ (with $s=p_-$ allowed).  Again, the story is similar for $p_- = 1$.

In the terminology of \cite{CreLog13}, the $\TripIrr{p_+,p_-}{\pm}$ are the \emph{typical} triplet modules, being direct sums of typical singlet modules.  The $\TripIrr{r,s}{\pm}$, with either $r \neq p_+$ or $s \neq p_-$, and the $\VirIrr{r,s}$ are then \emph{atypical} triplet modules.  We remark that the decompositions of the triplet simples into singlet simples suggest that the triplet algebra is just a simple current extension of the singlet algebra.  Indeed, the decomposition of $\TripIrr{1,1}{+}$ shows that the simple currents responsible for this conjectured extension are the $\SingIm{1,1;2n}$.  We will verify that this conjecture is consistent with our Verlinde formula computations in \secref{sec:SimpCurr}, though we will see that there are interesting subtleties which prevent the evidence from being conclusive.

\section{Characters and Modular Transformations for Singlet Models} \label{sec:SingFus}

This section details the derivation of the modular $\modS$-transformations of the characters of the simple $\SingAlg{p_+,p_-}$-modules.  The methodology follows the approach proposed in \cite{CreLog13} for general non-rational \cfts{}.  Specifically, the characters of the standard $\SingAlg{p_+,p_-}$-modules, simple and non-simple, are taken as a (topological) basis for a vector space which is shown to be preserved by the natural action of $\modS$.  This space then carries a representation of $\SLG{SL}{2;\ZZ}$ of uncountably-infinite dimension.  Resolutions are then derived for the atypical simple modules in terms of the non-simple standard modules and this gives expressions for the characters of the former as infinite alternating sums of the characters of the latter (the basis characters).  In this way, we arrive at $\modS$-transformations for all simple $\SingAlg{p_+,p_-}$-modules.

The computations are straightforward when $p_+$ or $p_-$ is 1.  However, the case where $p_+,p_- > 1$ is more interesting (as expected) because the ``$\modS$-matrix entries'', that describe the decomposition of the $\modS$-transformed character of $\VirIrr{r,s}$ into standard characters, are no longer functions of the parameters, but must instead be regarded as distributions.  Consequently, the same is true for the $\modS$-matrix entries of the vacuum module $\SingKer{1,1;0}$, leading to conceptual difficulties in applying the Verlinde formula.

\subsection{Characters of Feigin-Fuchs modules and their Modular Transformations}\label{sec:CharviaResol}

For any Virasoro module \(M\) at central charge \(c\), one defines its character to be the following 
power series in \(q=\exp(2\pi \ii \tau)\):
\begin{align}
  \ch{M}(\tau)=\traceover{M}\left(q^{L_0-c/24}\right).
\end{align}
For example, the characters of the Feigin-Fuchs modules \(\FF{\lambda}\) of \secref{sec:FF} are given by
\begin{align}\label{eq:FFqchar}
  \ch{\FF{\lambda}}(\tau)=\frac{q^{\Delta_\lambda-(1-3\alpha_0^2)/24}}{\prod_{i=1}^{\infty}\brac{1-q^i}}
  =\frac{q^{(\lambda-\alpha_0)^2/2}}{\eta(q)},
\end{align}
where \(\eta(q)\) is the Dedekind eta function.
As one can see from this formula, the Feigin-Fuchs modules
\(\FF{\lambda}\) and \(\FF{\alpha_0-\lambda}\) have identical characters. In
order to disambiguate these characters \cite{CreLog13}, we generalise them by adding an extra
formal variable \(z=\exp(2\pi \ii\zeta)\):
\begin{align}
  \ch{\FF{\lambda}}(\tau,\zeta)=\frac{q^{(\lambda-\alpha_0/2)^2/2}z^{\lambda-\alpha_0/2}}{\eta(q)}\,.
\end{align}
Here, one can think of $z$ as keeping track of the eigenvalue of the Heisenberg zero-mode (shifted by $-\alpha_0 / 2$).

The modular \(\modS\)-transformation of characters is the map\footnote{Technically, one needs an additional transformation variable to absorb the so-called automorphy factor.  We refer to \cite[Sec.~1.2]{CreLog13} for the detail (in the case $\alpha_0 = 0$).}
\begin{align}
  \ch{M}(\tau,\zeta)\mapsto \ch{M}(-\tfrac{1}{\tau},\tfrac{\zeta}{\tau})\,.
\end{align}
The characters of the Feigin-Fuchs modules satisfy the transformation formulae
\begin{align}
  \ch{\FF{\lambda}}(-\tfrac{1}{\tau},\tfrac{\zeta}{\tau})
  =\int_{\mathbb{R}} \Smat{\FF{\lambda}}{\FF{\rho}} \ch{\FF{\rho}}(\tau,\zeta)\dd \rho,
\end{align}
where the \emph{\(\modS\)-matrix coefficients} 
\(\tSmat{\FF{\lambda}}{\FF{\rho}}\) are given by
\begin{equation}\label{eq:StandardSCoeff}
\Smat{\FF{\lambda}}{\FF{\rho}}=\exp\sqbrac{-2\pi\ii(\lambda-\alpha_0/2)(\rho-\alpha_0/2)}.
\end{equation}
These coefficients follow from a straightforward gaussian integral, convergent
for $\Im \tau > 0$ hence $\abs{q} < 1$. In particular, for 
\(\lambda=\alpha_{r,s;n}\), the \(\modS\)-matrix coefficients specialise to
\begin{align}\label{eq:CornerSCoeff}
  \Smat{\FF{r,s;n}}{\FF{\rho}}=\ee^{\ii\pi r\alpha_+(\rho-\alpha_0/2)}\ee^{\ii\pi s\alpha_-(\rho-\alpha_0/2)}\ee^{-\ii\pi n\alpha(\rho-\alpha_0/2)}\,.
\end{align}

\subsection{Characters of Singlet Modules and their Modular Transformations}

With \eqnDref{eq:StandardSCoeff}{eq:CornerSCoeff}, we have determined the modular transformations of the standard $\SingAlg{p_+,p_-}$-modules.  The Felder complexes of \secref{sec:FF} and the exact sequences of
\secref{sec:Sing} now determine resolutions (or coresolutions or two-sided resolutions) of the singlet modules
\(\SingKer{r,s;n}^\bullet\), \(\SingIm{r,s;n}^\bullet\) and \(\VirIrr{r,s}\) in
terms of Feigin-Fuchs modules. These, in turn, allow us to calculate the characters
of the atypical singlet modules in terms of characters of the standard (Feigin-Fuchs) singlet modules.

It is important to note that the maps defining the Felder complexes and exact sequences are not Heisenberg algebra homomorphisms in general.  The interpretation of $z$ in Feigin-Fuchs characters as tracking the eigenvalue of the Heisenberg zero-mode therefore \emph{does not} lift to the character formulae for the atypical singlet modules that we shall derive.  There is no Heisenberg zero-mode in the singlet algebra, so the singlet characters should in fact be computed with $z=1$.  However, if one does this, one immediately encounters the problem that non-isomorphic singlet modules can have identical characters.  We will therefore keep $z$ as a formal variable in the singlet character formulae that follow.  Its function remains to naturally facilitate the distinguishing of characters of non-isomorphic modules, though it no longer appears to have any (obvious) interpretation in terms of eigenvalues of zero-modes.

\subsubsection{The \(\SingIm{r,s;n}^+\) modules}

As long as we avoid the non-exact parts of the Felder complexes
\eqref{eq:Felder}, we can use them to give (co)resolutions of the
\(\SingIm{r,s;n}^+\) in terms of Feigin-Fuchs modules, which in turn allow us
to derive character formulae and \(\modS\)-matrix coefficients.
When $n \geqslant 0$, we can iteratively splice the first exact sequence of \eqref{es:KerFFIm} with itself, using the isomorphisms \eqref{eq:Ker=Im}.  The result is a resolution for $\SingIm{r,s;n}^+$.  For $n<0$, the same method results instead in a coresolution.  We therefore obtain, for \(1\leqslant r\leqslant p_+-1\) and \(1\leqslant s\leqslant p_-\), the following (co)resolutions of the $\SingIm{r,s;n}^+$:
\begin{subequations}
\begin{align}\label{eq:ResImpa}
  \cdots 
  \lra \FF{p_+ - r,s;n-3} \lra
    \FF{r,s;n-2} \lra \FF{p_+ - r,s;n-1}
    \lra \SingIm{r,s;n}^+ \lra 0 \qquad &\text{\((n \leqslant 0\)),}\\\label{eq:ResImpb}
    0 \lra \SingIm{r,s;n}^+ \lra
    \FF{r,s;n} \lra \FF{p_+ - r,s;n+1}
    \lra \FF{r,s;n+2} \lra \cdots \qquad &\text{(\(n \geqslant 1\)).}
\end{align}
\end{subequations}
Note that both of these sequences will be exact for all \(n\in\mathbb{Z}\) if and only if
\(s=p_-\).
From these (co)resolutions, we can read off the character formulae
\begin{align} \label{ch:Im+}
  \ch{\SingIm{r,s;n}^+}=
    \begin{cases}
      \displaystyle\sum_{k\geqslant 0}\left(\ch{\FF{p_+-r,s;n-2k-1}}-\ch{\FF{r,s;n-2k-2}}\right)&\text{if \(n\leqslant 0\),}\\
      \displaystyle\sum_{k\geqslant 0}\left(\ch{\FF{r,s;n+2k}}-\ch{\FF{p_+-r,s;n+2k+1}}\right)&\text{if \(n\geqslant 1\)}
    \end{cases}
\end{align}
and the \(\modS\)-matrix coefficients are obtained by adding and subtracting the \(\modS\)-matrix entries \(\tSmat{\FF{p_+ - r,s;n-2k-1}}{\FF{\rho}}\) and \(\tSmat{\FF{r,s;n-2k-2}}{\FF{\rho}}\), for $n \leqslant 0$, or \(\tSmat{\FF{r,s;n+2k}}{\FF{\rho}}\) and \(\tSmat{\FF{p_+ - r,s;n+2k+1}}{\FF{\rho}}\), for $n \geqslant 1$:
\begin{align}
  \Smat{\SingIm{r,s;n}^+}{\FF{\rho}}=
    \begin{cases}
      -2\ii\sin[\pi r\alpha_+(v-\alpha_0/2)]\ee^{\ii\pi
        s\alpha_-}\ee^{-\ii\pi(n-2)\alpha(\rho-\alpha_0/2)}
      \displaystyle\sum_{k\geqslant 0}\ee^{2\pi\ii k\alpha(\rho-\alpha_0/2)}
      & \text{if \(n\leqslant 0\),}\\
      +2i\sin[\pi r\alpha_+(v-\alpha_0/2)]\ee^{\ii\pi
        s\alpha_-}\ee^{-\ii\pi n\alpha(\rho-\alpha_0/2)}
      \displaystyle\sum_{k\geqslant 0}\ee^{-2\pi\ii k\alpha(\rho-\alpha_0/2)}
      & \text{if \(n\geqslant 1\).}
    \end{cases}
\end{align}
The infinite sums in this formula are geometric
series at the boundaries of their radius of convergence. Nevertheless, we will
replace these geometric series by their analytic continuations\footnote{We
  refer to \cite{CreFal13} for an explicit example of how to regularise these
  sums in the case that \(p_+\) or \(p_-\) equal $1$. It is not clear to us if this regularisation extends to \(p_+\) and \(p_-\) greater than $1$.}
\begin{align}\label{eq:GeomSeries}
  \sum_{k\geq0} x^{+2k} \longmapsto \frac{-x^{-1}}{x-x^{-1}}, \qquad 
  \sum_{k\geq0} x^{-2k} \longmapsto \frac{x}{x-x^{-1}}.
\end{align}
With these replacements,
the \(\modS\)-matrix coefficients simplify to the common form (for all $n$)
\begin{align}\label{eq:pSCoeff}
  \Smat{\SingIm{r,s;n}^+}{\FF{\rho}}=
  \ee^{\ii\pi s\alpha_-(\rho-\alpha_0/2)}\ee^{-\ii\pi(n-1)\alpha(\rho-\alpha_0/2)}
  \frac{\sin[\pi r\alpha_+(\rho-\alpha_0/2)]}{\sin[\pi
    p_+\alpha_+(\rho-\alpha_0/2)]} \qquad \text{(\(r \neq p_+\)),}
\end{align}
where we have used the identity \(\alpha=p_+\alpha_+\) in the denominator.

\subsubsection{The \(\SingIm{r,s;n}^-\) modules}

The derivation of (co)resolutions, character formulae and \(\modS\)-matrix
coefficients for the singlet modules \(\SingIm{r,s;n}^-\) is completely
analogous to the derivation for the \(\SingIm{r,s;n}^+\).

For \(1\leqslant r\leqslant p_+\) and \(1\leqslant s\leqslant p_--1\), the following sequences are exact
and define (co)resolutions of \(\SingIm{r,s;n}^-\):
\begin{subequations} \label{eq:ResImm}
\begin{align}\label{eq:ResImma}
  \cdots 
  \lra \FF{r,p_--s;n+3} \lra
    \FF{r,s;n+2} \lra \FF{r,p_--s;n+1}
    \lra \SingIm{r,s;n}^- \lra 0 \qquad &\text{(\(n \geqslant 0\)),}\\\label{eq:ResImmb}
    0 \lra \SingIm{r,s;n}^- \lra
    \FF{r,s;n} \lra \FF{r,p_--s;n-1}
    \lra \FF{r,s;n-2} \lra \cdots \qquad &\text{(\(n \leqslant -1\)).}
\end{align}
\end{subequations}
These (co)resolutions lead to the character formulae
\begin{align} \label{ch:Im-}
  \ch{\SingIm{r,s;n}^-}=
    \begin{cases}
      \displaystyle\sum_{k\geqslant 0}\left(\ch{\FF{r,p_--s;n+2k+1}}-\ch{\FF{r,s;n+2k+2}}\right)& \text{if \(n\geqslant 0\),}\\
      \displaystyle\sum_{k\geqslant 0}\left(\ch{\FF{r,s;n-2k}}-\ch{\FF{r,p_--s;n-2k-1}}\right)& \text{if \(n\leqslant -1\),}
    \end{cases}
\end{align}
which in turn imply the \(\modS\)-matrix coefficients
\begin{align}\label{eq:mSCoeff}
  \Smat{\SingIm{r,s;n}^-}{\FF{\rho}}=
  \ee^{\ii\pi r\alpha_+(\rho-\alpha_0/2)}\ee^{-\ii\pi(n+1)\alpha(\rho-\alpha_0/2)}
  \frac{\sin[\pi s\alpha_-(\rho-\alpha_0/2)]}{\sin[\pi
    p_-\alpha_-(\rho-\alpha_0/2)]}\qquad \text{(\(s \neq p_-\)),}
\end{align}
for general $n$.  Here, we have used the identity \(\alpha=-p_-\alpha_-\) in the denominator as well as 
the analytic continuations \eqref{eq:GeomSeries}.

\subsubsection{The \(\SingIm{r,s;n}\) modules}
For \(1\leqslant r\leqslant p_+-1\) and \(1\leqslant s\leqslant p_--1\), the singlet modules
\(\SingIm{r,s;n}\) can be resolved by iteratively splicing the exact sequences
\eqref{es:ImImIm} to obtain
\begin{subequations}
\begin{align}\label{eq:ResIma}
  \cdots 
  \lra \SingIm{r,p_--s;n+3}^+ \lra
    \SingIm{r,s;n+2}^+ \lra \SingIm{r,p_--s;n+1}^+
    \lra \SingIm{r,s;n} \lra 0, \\\label{eq:ResImb}
  \cdots 
  \lra \SingIm{p_+-r,s;n-3}^- \lra
    \SingIm{r,s;n-2}^- \lra \SingIm{p_+-r,s;n-1}^-
    \lra \SingIm{r,s;n} \lra 0.
\end{align}
\end{subequations}
We therefore arrive at two seemingly different character formulae:
\begin{subequations}
\begin{align}
  \ch{\SingIm{r,s;n}}&=\sum_{k\geqslant 0}\left(\ch{\SingIm{r,p_--s;n+2k+1}^+}-\ch{\SingIm{r,s;n+2k+2}^+}\right)\\
  &=\sum_{k\geqslant 0}\left(\ch{\SingIm{p_+-r,s;n-2k-1}^-}-\ch{\SingIm{r,s;n-2k-2}^-}\right).
\end{align}
\end{subequations}
However, the resulting \(\modS\)-matrix coefficients are identical:
\begin{align}\label{eq:IntSCoeff}
  \Smat{\SingIm{r,s;n}}{\FF{\rho}}=\frac{\sin[\pi
    r\alpha_+(\rho-\alpha_0/2)]}{\sin[\pi p_+\alpha_+(\rho-\alpha_0/2)]}
  \frac{\sin[\pi s\alpha_-(\rho-\alpha_0/2)]}{\sin[\pi
    p_-\alpha_-(\rho-\alpha_0/2)]}
  \ee^{-\ii\pi n\alpha(\rho-\alpha_0/2)}.
\end{align}
One can also coresolve the $\SingIm{r,s;n}$ in terms of the $\SingIm{r,s;n}^{\pm}$ with the same result.

\subsubsection{The $\VirIrr{r,s}$ Modules}

As noted in \secref{sec:FF}, the cohomology of the Felder
complexes \eqref{eq:Felder} is trivial everywhere except when \(1\leqslant
r\leqslant p_+-1\), \(1\leqslant s\leqslant p_--1\) and $n=0$, where it is the simple Virasoro module
\(\VirIrr{r,s}\). 
By the Euler-Poincar\'{e} principle, the Felder complex \eqref{eq:Felder+}
implies that the character of the simple singlet module \(\VirIrr{r,s}\) is
given by
\begin{subequations} \label{ch:Lrs}
\begin{align} \label{ch:Lrs+}
  \ch{\VirIrr{r,s}}^+=\sum_{n\in\mathbb{Z}}\left(\ch{\FF{r,s;2n}}-\ch{\FF{p_+-r,s;2n+1}}\right),
\end{align}
whereas the Felder complex \eqref{eq:Felder-}
gives the character of \(\VirIrr{r,s}\) as
\begin{align} \label{ch:Lrs-}
  \ch{\VirIrr{r,s}}^-=\sum_{n\in\mathbb{Z}}\left(\ch{\FF{r,s;2n}}-\ch{\FF{r,p_--s;2n+1}}\right).
\end{align}
\end{subequations}
It is not clear that these coincide.  However, we remark that we should only expect that these character formulae coincide once we remove the $z$-dependence.  Recall that we included $z$, somewhat artificially, in our definition of singlet characters so as to be able to distinguish simple modules that would otherwise have identical characters. Setting $z=1$, it is easy to check that the right-hand sides of \eqref{ch:Lrs} indeed coincide formally as a consequence of \eqref{eq:FFqchar} and the identity $\alpha_{p_+-r,s;2n+1} - \alpha_0/2 = -\brac{\alpha_{r,p_--s;-2n-1} - \alpha_0/2}$.

Now, unlike the character formulae for the $\SingIm{r,s;n}$ considered above, the character formulae \eqref{ch:Lrs} yield \emph{different} \(\modS\)-matrix coefficients:
\begin{subequations}
\begin{align}
  \Smat{\VirIrr{r,s}}{\FF{\rho}}^+&=2\ii\sin[\pi r\alpha_+(\rho-\alpha_0/2)]\ee^{\ii\pi
    s\alpha_-(\rho-\alpha_0/2)}
  \sum_{n\in\mathbb{Z}}\ee^{-2\pi\ii
    n\alpha(\rho-\alpha_0/2)},\label{eq:VirSCoeffp}\\\label{eq:VirSCoeffm}
  \Smat{\VirIrr{r,s}}{\FF{\rho}}^-&=2\ii\sin[\pi s\alpha_-(\rho-\alpha_0/2)]\ee^{\ii\pi
    r\alpha_+(\rho-\alpha_0/2)}
  \sum_{n\in\mathbb{Z}}\ee^{-2\pi\ii
    n\alpha(\rho-\alpha_0/2)}.
\end{align}
\end{subequations}
The superscript ``$\pm$'' serves to remind us which Felder complex was used in the derivation.  We note that the sums in these formulae do not define functions but must be interpreted as distributions (see \secref{sec:Decoupling}).  We note in addition that the character formulae \eqref{ch:Lrs} also do not appear to respect the isomorphism $\VirIrr{r,s} \cong \VirIrr{p_+ - r,p_- - s}$.  Again, the disambiguation variable $z$ is to blame.  We shall see shortly that this non-uniqueness problem solves itself rather naturally in the setting of the Verlinde algebra, though there is of course a price that has to be paid.
  
\subsubsection{The \(\SingKer{r,s;0}\) modules}

For \(1\leqslant r\leqslant p_+-1\) and \(1\leqslant s\leqslant p_--1\), the exact sequences
\eqref{es:KerFFIm} and the formulae \eqref{ch:Lrs} give two distinct character formulae for the \(\SingKer{r,s;0}\) singlet modules:
\begin{align}
  \ch{\SingKer{r,s;0}}^\pm=\ch{\VirIrr{r,s}}^\pm+\ch{\SingIm{r,s;0}}.
\end{align}
These in turn yield two distinct \(\modS\)-matrix coefficients:
\begin{subequations} \label{eq:KSCoeff}
\begin{align}
  \Smat{\SingKer{r,s;0}}{\FF{\rho}}^+&=2\ii\sin[\pi r\alpha_+(\rho-\alpha_0/2)]\ee^{\ii\pi
    s\alpha_-(\rho-\alpha_0/2)}
  \sum_{n\in\mathbb{Z}}\ee^{-2\pi\ii
    n\alpha(\rho-\alpha_0/2)} \nonumber \\
  &\mspace{20mu}+\frac{\sin[\pi
    r\alpha_+(\rho-\alpha_0/2)]}{\sin[\pi p_+\alpha_+(\rho-\alpha_0/2)]}
  \frac{\sin[\pi s\alpha_-(\rho-\alpha_0/2)]}{\sin[\pi
    p_-\alpha_-(\rho-\alpha_0/2)]}
  \ee^{-\ii\pi n\alpha(\rho-\alpha_0/2)},\\
  \Smat{\SingKer{r,s;0}}{\FF{\rho}}^-&=2\ii\sin[\pi s\alpha_-(\rho-\alpha_0/2)]\ee^{\ii\pi
    r\alpha_+(\rho-\alpha_0/2)}
  \sum_{n\in\mathbb{Z}}\ee^{-2\pi\ii
    n\alpha(\rho-\alpha_0/2)} \nonumber \\
  &\mspace{20mu}+\frac{\sin[\pi
    r\alpha_+(\rho-\alpha_0/2)]}{\sin[\pi p_+\alpha_+(\rho-\alpha_0/2)]}
  \frac{\sin[\pi s\alpha_-(\rho-\alpha_0/2)]}{\sin[\pi
    p_-\alpha_-(\rho-\alpha_0/2)]}
  \ee^{-\ii\pi n\alpha(\rho-\alpha_0/2)}.
\end{align}
\end{subequations}
We remark that the vacuum module is $\SingKer{1,1;0}$, so this non-uniqueness
is a potential problem for the Verlinde computations below.  This is on top of the problem that one will seemingly have to make sense of dividing by a distribution in order to apply the Verlinde formula.  Again, we will resolve these issues in \secref{sec:Decoupling}.  Note however that the above formulae are not valid for $\SingKer{1,1;0}$ when $p_+ = 1$ or $p_- = 1$.  In these cases, $\SingKer{1,1;0} \cong \SingIm{1,1;0}$ and the vacuum $\modS$-matrix coefficients are instead given in \eqnref{eq:IntSCoeff}.

\subsection{Addendum}

At this point, the reader might object that the character formulae for the $\SingIm{r,s;n}^{\pm}$ have been derived from (co)resolutions which avoid the part where the Felder complexes \eqref{eq:Felder} fail to be exact.  Is it possible that we will obtain different formulae if we (co)resolve so as to cross the non-exact piece of the complexes?  Now that we have the characters of the $\VirIrr{r,s}$, derived from each Felder complex, it is easy to attend to this objection.  We will consider the resolution \eqref{eq:ResImpa} for $\SingIm{r,s;n}^+$ with $r \neq p_+$ and $n \geqslant 1$, the analysis in the other cases being almost identical.

First, the splicing of the exact sequences \eqref{eq:Ker=Im} stops when the third label ($n$) of the modules drops to $0$ because $\SingKer{r,s;0}^+$ and $\SingIm{r,s;0}^+$ are not isomorphic.  Instead, we arrive at the long exact sequence
\begin{equation}
0 \lra \SingKer{*,s;0}^+ \lra \FF{*,s;0} \lra \cdots \lra \FF{r,s;n-2} \lra \FF{p_+ - r,s;n-1} \lra \SingIm{r,s;n}^+ \lra 0,
\end{equation}
where the subscript ``$*$'' stands for $r$, if $n$ is even, and $p_+ - r$, if $n$ is odd.  Because of the exact sequence \eqref{es:ImKerVir}, continuing the splicing to obtain \eqref{eq:ResImpa} does not yield a resolution of $\SingIm{r,s;n}^+$ (for $n \geqslant 1$) because the sequence has non-trivial cohomology $\VirIrr{r,s}$ when the third label is $0$.  No matter --- Euler-Poincar\'{e} says that the character formula \eqref{ch:Im+} derived earlier for the $\SingIm{r,s;n}^+$ with $n \leqslant 0$ should be corrected, for $n>0$, by $(-1)^n \ch{\VirIrr{*,s}}$:
\begin{equation}
\ch{\SingIm{r,s;n}^+} = 
\begin{cases}
\displaystyle\sum_{k\geqslant 0}\left(\ch{\FF{p_+-r,s;n-2k-1}}-\ch{\FF{r,s;n-2k-2}}\right) + \ch{\VirIrr{r,s}} & \text{if \(n \geqslant 1\) is even,} \\
\displaystyle\sum_{k\geqslant 0}\left(\ch{\FF{p_+-r,s;n-2k-1}}-\ch{\FF{r,s;n-2k-2}}\right) - \ch{\VirIrr{p_+-r,s}} & \text{if \(n \geqslant 1\) is odd.}
\end{cases}
\end{equation}
Because the $\SingKer{r,s;0}^+$ and $\SingIm{r,s;0}^+$ are defined using the Felder complex \eqref{eq:Felder+}, we must use the character formula \eqref{ch:Lrs+} for the $\VirIrr{r,s}$ that was derived from this complex.  When we substitute this into the above character formulae, we find that the result precisely reproduces \eqref{ch:Im+} for $n \geqslant 1$.  In summary, the character formula does not depend upon whether we use a (not quite exact) resolution or a coresolution.

\section{The Verlinde Ring for Singlet Models} \label{sec:Verlinde}

Everything is now in place to use the obvious continuum analogue of the Verlinde formula to compute ``fusion coefficients'' for the singlet theories.  Roughly speaking, this will define a ring structure on an abelian group generated by (some of) the simple characters.  We will refer to this ring as the (singlet) \emph{Verlinde ring}, denoting it by $\SingVerRing{p_+,p_-}$.  As we will see, the computations are relatively straightforward, although there is the issue of the distributional nature of the vacuum $\modS$-matrix entries to surmount.
However, the interpretation of the computations is not:  For $\SingAlg{1,1}$, we do indeed recover the fusion coefficients because of the semisimplicity of the module category.  The Verlinde ring $\SingVerRing{1,1}$ is the fusion ring of the Heisenberg algebra.  For $\SingAlg{1,p_-}$ and $\SingAlg{p_+,1}$, non-semisimple modules exist, so the best we can expect is that the Verlinde ring will coincide with the \emph{Grothendieck ring} of fusion.  This, in turn, requires that fusion defines an exact functor on the singlet modules which we expect to be true (it is true for the triplet algebras $\TripAlg{1,p_-}$ and $\TripAlg{p_+,1}$ \cite{TsuTen12}).  However, fusion is not expected to be exact for $p_+,p_- > 1$ and, indeed, non-exactness has been demonstrated for the corresponding triplet algebras \cite{GabFus09}.  So, we can only expect that the Verlinde ring may be identified with a quotient of the Grothendieck group upon which fusion restricts to an exact product.  Happily, the appropriate quotient is naturally determined from simple considerations:  It is obtained by setting the $\VirIrr{r,s}$ to $0$ --- see \secref{sec:Decoupling}.

Before we start calculating the structure constants of the Verlinde ring, it is convenient to extend the range of the indices of the \(\SingIm{r,s;n}\) to include \(r=p_+\) and \(s=p_-\). Let \(1\leqslant
r\leqslant p_+-1\) and \(1\leqslant s\leqslant p_--1\) and then define
\begin{align} \label{eq:AllAreIm}
  \SingIm{p_+,s;n}:=\SingIm{p_+,s;n}^-,\qquad\SingIm{r,p_-;n}:=\SingIm{r,p_-;n}^+,\qquad
  \SingIm{p_+,p_-;n}:=\FF{p_+,p_-;n}.
\end{align}
For \(1\leqslant r\leqslant p_+\) and \(1\leqslant s\leqslant p_-\), the \(\modS\)-matrix coefficients
\eqref{eq:CornerSCoeff}, \eqref{eq:pSCoeff}, \eqref{eq:mSCoeff} and
\eqref{eq:IntSCoeff} can now be compactly written in the unified form
\begin{align}\label{eq:ImSCoeff}
  \Smat{\SingIm{r,s;n}}{\FF{\rho}}=\frac{\sin[\pi
    r\alpha_+(\rho-\alpha_0/2)]}{\sin[\pi p_+\alpha_+(\rho-\alpha_0/2)]}
  \frac{\sin[\pi s\alpha_-(\rho-\alpha_0/2)]}{\sin[\pi
    p_-\alpha_-(\rho-\alpha_0/2)]}
  \ee^{-\ii\pi n\alpha(\rho-\alpha_0/2)}.
\end{align}

\subsection{The Decoupling of the \(\VirIrr{r,s}\)} \label{sec:Decoupling}

For two singlet modules \(M\) and \(N\), the Verlinde formula for ``fusion products'' states that there is a product (on certain equivalence classes of modules) given by
\begin{align} \label{eq:Verlinde}
  \Gr{M}\fuse \Gr{N} =\int_{\mathbb{R}}\fuscoeff{M}{N}{\nu} \Gr{\FF{\nu}}\: \dd\nu.
\end{align}
The ``fusion coefficents'' \(\fuscoeff{M}{N}{\nu}\) are defined in terms of the \(\modS\)-matrix coefficients by
\begin{align} \label{eq:DefFusCoeff}
  \fuscoeff{M}{N}{\nu}=\int_{\mathbb{R}}\frac{\tSmat{M}{\FF{\rho}}\tSmat{N}{\FF{\rho}}\tSmat{\FF{\nu}}{\FF{\rho}}^*}
  {\tSmat{\SingKer{1,1;0}}{\FF{\rho}}}\:\dd\rho,
\end{align}
where ``$*$'' denotes complex conjugation.  Consider the denominator of the above integrand.  According to \eqnref{eq:KSCoeff}, this \(\modS\)-matrix coefficient has two possible forms:
\begin{align}
  \Smat{\SingKer{1,1;0}}{\FF{\rho}}^{\pm}&=\Smat{\VirIrr{1,1}}{\FF{\rho}}^\pm+\Smat{\SingIm{1,1;0}}{\FF{\rho}}\nonumber\\
  &=\Smat{\SingIm{1,1;0}}{\FF{\rho}}\left(1+\frac{\tSmat{\VirIrr{1,1}}{\FF{\rho}}^\pm}{\tSmat{\SingIm{1,1;0}}{\FF{\rho}}}\right)\,.
\end{align}
The quotient of the \(\modS\)-matrix coefficients of \(\VirIrr{1,1}\) and
\(\SingIm{1,1;0}\) can be evaluated using \eqref{eq:VirSCoeffp} and \eqref{eq:ImSCoeff}:
\begin{align} \label{eq:ThisIsZero}
  &\frac{\tSmat{\VirIrr{1,1}}{\FF{\rho}}^+}{\tSmat{\SingIm{1,1;0}}{\FF{\rho}}}=
  \frac{\sin[\pi
    p_+\alpha_+(\rho-\alpha_0/2)]}{\sin[\pi\alpha_+(\rho-\alpha_0/2)]}
  \frac{\sin[\pi
    p_-\alpha_-(\rho-\alpha_0/2)]}{\sin[\pi\alpha_-(\rho-\alpha_0/2)]}\nonumber\\
  &\mspace{135mu}\cdot 2\ii\sin[\pi \alpha_+(\rho-\alpha_0/2)]\ee^{\ii\pi
    \alpha_-(\rho-\alpha_0/2)}
  \sum_{k\in\mathbb{Z}}\ee^{-2\pi\ii\alpha(\rho-\alpha_0/2)k}\nonumber\\
  &= \ee^{\ii\pi\alpha_-(\rho-\alpha_0/2)}
  \frac{\sin[\pi p_-\alpha_-(\rho-\alpha_0/2)]}{\sin[\pi\alpha_-(\rho-\alpha_0/2)]}
  \cdot2\ii\sin[\pi \alpha(\rho-\alpha_0/2)]
  \sum_{k\in\mathbb{Z}}\ee^{-2\pi\ii\alpha(\rho-\alpha_0/2)k}\,.
\end{align}
The product of the last two factors on the right-hand-side can be identified with zero because
\begin{align}
  2\ii\sin[\pi\alpha(\rho-\alpha_0/2)]\sum_{k\in\mathbb{Z}}\ee^{2\pi\ii\alpha(\rho-\alpha_0/2)k}
  &=\ee^{\ii\pi\alpha(\rho-\alpha_0/2)}(1-e^{-2\pi\ii\alpha(\rho-\alpha_0)})
  \sum_{k\in\mathbb{Z}}\ee^{-2\pi\ii\alpha(\rho-\alpha_0/2)k}\nonumber\\
  &=\ee^{\ii\pi\alpha(\rho-\alpha_0/2)}
  \sum_{k\in\mathbb{Z}}(\ee^{-2\pi\ii\alpha(\rho-\alpha_0/2)k}-\ee^{-2\pi\ii\alpha(\rho-\alpha_0/2)(k+1)})\,.
\end{align}
A similar calculation gives the same conclusion for \(\tSmat{\VirIrr{1,1}}{\FF{\rho}}^-\). 
We may therefore replace the vacuum
\(\modS\)-matrix coefficient \(\tSmat{\SingKer{1,1;0}}{\FF{\rho}}^\pm\) in the Verlinde formula by 
\(\tSmat{\SingIm{1,1;0}}{\FF{\rho}}\), the \(\modS\)-matrix coefficient of its simple submodule
\(\SingIm{1,1;0}\).  In this way, the non-uniqueness of the vacuum \(\modS\)-matrix coefficients is neatly bypassed.

We remark that the \(\modS\)-matrix coefficients \(\tSmat{\VirIrr{r,s}}{\FF{\rho}}^\pm\), and hence the \(\tSmat{\SingKer{r,s;0}}{\FF{\rho}}^\pm\) as well, can only be understood as distributions:
\begin{subequations}
\begin{align}
\Smat{\VirIrr{r,s}}{\FF{\rho}}^+ &= +\frac{1}{\alpha} \sum_{k \in \ZZ} 2 \ii \sin \frac{\pi k r}{p_+} \ee^{-\ii \pi k s/p_-} \func{\delta}{\rho - \alpha_0 / 2 - k / \alpha}, \\
\Smat{\VirIrr{r,s}}{\FF{\rho}}^- &= -\frac{1}{\alpha} \sum_{k \in \ZZ} 2 \ii \sin \frac{\pi k s}{p_-} \ee^{+\ii \pi k r/p_+} \func{\delta}{\rho - \alpha_0 / 2 - k / \alpha}.
\end{align}
\end{subequations}
From this point of view, the vanishing of the quotient \eqref{eq:ThisIsZero} and its ``$-$'' version is manifest because the factors $\sin [\pi p_{\pm} \alpha_{\pm} (\rho - \alpha_0 / 2)]$ may be replaced by $\pm \sin [k \pi] = 0$ when brought into the sum over $k$, so the coefficient of each delta function is zero.

We next argue that at the level of the Verlinde products, the minimal model
modules \(\VirIrr{r,s}\) decouple, meaning that for any singlet module
\(N\), we have \(\fuscoeff{\VirIrr{r,s}}{N}{\rho}=0\). This follows from an obvious generalisation of the argument above for $(r,s)=(1,1)$ to the quotients
\[
  \frac{\tSmat{\VirIrr{r,s}}{\FF{\rho}}^\pm}{\tSmat{\SingIm{1,1;0}}{\FF{\rho}}}.
\]
It therefore follows from \eqref{eq:DefFusCoeff} and the replacement \(\tSmat{\SingKer{1,1;0}}{\FF{\rho}}^\pm \ra \tSmat{\SingIm{1,1;0}}{\FF{\rho}}\) that \(\fuscoeff{\VirIrr{r,s}}{N}{\rho}=0\).  Summarising, we have seen that the alarming non-uniqueness and distributional nature of the \(\modS\)-matrix coefficients that we observed for the $\VirIrr{r,s}$ is not at all troublesome for Verlinde computations as these modules decouple completely.  In other words, the $\VirIrr{r,s}$ may be naturally identified with zero in the Verlinde ring:  $\Gr{\VirIrr{r,s}} = 0$.

\subsection{Verlinde Products}

We are now ready to compute the ``Verlinde products'' defined by the Verlinde formula.  First, we will compute the product \(\Gr{\FF{\lambda}}\fuse \Gr{\FF{\mu}}\) of standard singlet module characters for $\lambda, \mu \in \RR$.\footnote{In what follows, we will drop the ``$\chmap$'' from the character $\ch{M}$ of a module $M$ for brevity and to make contact with \eqnref{eq:Verlinde}.}  The Verlinde coefficient we seek is
\begin{align}
\fuscoeff{\FF{\lambda}}{\FF{\mu}}{\nu} &= \int_{\RR} \frac{\tSmat{\FF{\lambda}}{\FF{\rho}} \tSmat{\FF{\mu}}{\FF{\rho}} \tSmat{\FF{\nu}}{\FF{\rho}}^*}{\tSmat{\SingIm{1,1;0}}{\FF{\rho}}}\:\dd\rho \nonumber \\
&= \int_{\RR} \frac{\sin[\pi p_+ \alpha_+ (\rho - \alpha_0 / 2)]}{\sin[\pi \alpha_+ (\rho - \alpha_0/2)]} \frac{\sin[\pi p_- \alpha_- (\rho - \alpha_0/2)]}{\sin[\pi \alpha_- (\rho - \alpha_0/2)]} \ee^{-2 \pi \ii (\lambda + \mu - \nu - \alpha_0 / 2) (\rho - \alpha_0/2)} \: \dd \rho
\end{align}
which, by means of the trigonometric identity
\begin{equation} \label{eq:TrigId}
  \frac{\sin[px]}{\sin[x]}=\frac{\ee^{\ii px}-\ee^{-\ii px}}{\ee^{\ii x}-\ee^{-\ii x}}
  =\sum_{j=0}^{p-1}\ee^{\ii (p-1-2j) x},
\end{equation}
simplifies to
\begin{align}
\fuscoeff{\FF{\lambda}}{\FF{\mu}}{\nu} &= \sum_{j_+ = 0}^{p_+ - 1} \sum_{j_- = 0}^{p_- - 1} \int_{\RR} \ee^{-2 \pi \ii (\lambda + \mu - \nu + j_+ \alpha_+ + j_- \alpha_-) (\rho - \alpha_0 / 2)} \: \dd \rho \nonumber \\
&= \sum_{j_+ = 0}^{p_+ - 1} \sum_{j_- = 0}^{p_- - 1} \func{\delta}{\nu - \lambda - \mu - j_+ \alpha_+ - j_- \alpha_-}.
\end{align}
The Verlinde formula \eqref{eq:Verlinde} therefore yields the following product:
\begin{align}\label{eq:FFTimesFF}
\Gr{\FF{\lambda}} \fuse \Gr{\FF{\mu}} = \sum_{j_+ = 0}^{p_+ - 1} \sum_{j_- = 0}^{p_- - 1} \Gr{\FF{\lambda + \mu + j_+ \alpha_+ + j_- \alpha_-}}.
\end{align}

We next turn to the product \(\Gr{\SingIm{r,s;n}}\fuse \Gr{\FF{\mu}}\) for \(1\leqslant r\leqslant p_+\), \(1\leqslant s\leqslant p_+\), \(n\in\mathbb{Z}\) and \(\mu\in\mathbb{R}\). The remaining Verlinde products will then be calculated by applying this product to the appropriate (co)resolutions.  This time, the Verlinde coefficient to be computed is
\begin{align}
  \fuscoeff{\SingIm{r,s;n}}{\FF{\mu}}{\nu}&=
  \int_{\mathbb{R}}
  \frac{\sin[\pi
    r\alpha_+(\rho-\alpha_0/2)]}{\sin[\pi \alpha_+(\rho-\alpha_0/2)]}
  \frac{\sin[\pi s\alpha_-(\rho-\alpha_0/2)]}{\sin[\pi
    \alpha_-(\rho-\alpha_0/2)]}
  \ee^{-\ii\pi n\alpha(\rho-\alpha_0/2)}\ee^{-2\pi\ii(\mu-\nu)(\rho-\alpha_0/2)}\:\dd\rho \nonumber \\
  &=\sum_{j_+=0}^{r-1}\sum_{j_-=0}^{s-1}\func{\delta}{\nu - \alpha_{r,s;n}-\mu-j_+\alpha_+-j_-\alpha_-}.
\end{align}
The Verlinde formula therefore gives
\begin{align}\label{eq:ImTimesFF}
  \Gr{\SingIm{r,s;n}}\fuse\Gr{\FF{\mu}}=\sum_{j_+=0}^{r-1}\sum_{j_-=0}^{s-1}\Gr{\FF{\alpha_{r-2j_+,s-2j_-;n}+\mu}}.
\end{align}

As the Verlinde formula is defined entirely in terms of characters of modules,
it cannot differentiate between an indecomposable module and the
direct sum of its simple composition factors. We therefore obtain the following identifications, for 
\(1\leqslant r\leqslant p_+-1\) and \(1\leqslant s\leqslant p_--1\), from Equations~\eqref{eq:Ker=Im}, \eqref{es:KerFFIm} and \eqref{es:ImImIm}:
\begin{equation} \label{eq:MoreIdents}
\begin{aligned}
  \Gr{\SingIm{r,s;n}^+}&=\Gr{\SingIm{r,s;n}}+\Gr{\SingIm{r,p_--s;n-1}}, \\
  \Gr{\SingIm{r,s;n}^-}&=\Gr{\SingIm{r,s;n}}+\Gr{\SingIm{p_+-r,s;n+1}},
\end{aligned}
\qquad
\begin{aligned}
  \Gr{\FF{p_+,s;n}}&=\Gr{\SingIm{p_+,s;n}}+\Gr{\SingIm{p_+,p_--s;n-1}}, \\
  \Gr{\FF{r,p_-;n}}&=\Gr{\SingIm{r,p_-;n}}+\Gr{\SingIm{p_+-r,p_-;n+1}}.
\end{aligned}
\end{equation}
Here, we recall the definitions \eqref{eq:AllAreIm} that we have already made for $r=p_+$ or $s=p_-$.
By applying the Verlinde product \eqref{eq:ImTimesFF} to the (co)resolutions
\eqref{eq:ResImpa}, \eqref{eq:ResImpb}, \eqref{eq:ResImma},
\eqref{eq:ResImmb}, \eqref{eq:ResIma}
and \eqref{eq:ResImb},
we obtain the following products involving the \(\SingIm{1,1;m}\),
\(\SingIm{2,1;0}\) and \(\SingIm{1,2;0}\):
\begin{subequations} \label{eq:GeneratorFusion}
\begin{align}
  \Gr{\SingIm{1,1;m}}\fuse \Gr{\SingIm{r,s;n}}&=\Gr{\SingIm{r,s;m+n}}, \label{eq:SimpCurrFusion} \\
  \Gr{\SingIm{2,1;0}}\fuse \Gr{\SingIm{r,s;n}}&=
    \begin{cases}
      \Gr{\SingIm{2,s;n}} & \text{if \(r=1\),}\\
      \Gr{\SingIm{r-1,s;n}}+\Gr{\SingIm{r+1,s;n}} & \text{if \(1< r< p_+\),}\\
      \Gr{\SingIm{1,s;n-1}}+2\:\Gr{\SingIm{p_+-1,s;n}}+\Gr{\SingIm{1,s;n+1}} & \text{if \(r=p_+\),}
    \end{cases}
  \\
  \Gr{\SingIm{1,2;0}}\fuse \Gr{\SingIm{r,s;n}}&=
    \begin{cases}
      \Gr{\SingIm{r,2;n}} & \text{if \(s=1\),}\\
      \Gr{\SingIm{r,s-1;n}}+\Gr{\SingIm{r,s+1;n}} & \text{if \(1< s< p_-\),}\\
      \Gr{\SingIm{r,1;n-1}}+2\:\Gr{\SingIm{r,p_--1;n}}+\Gr{\SingIm{r,1;n+1}} & \text{if \(s=p_-\).}
    \end{cases}
\end{align}
\end{subequations}
We note for future use that the linear $\ZZ$-span of the $\Gr{\SingIm{r,s;n}}$ is closed under the Verlinde product.

To illustrate how to apply \eqref{eq:ImTimesFF} to (co)resolutions, we present the details of the derivation of the product $\Gr{\SingIm{2,1;0}}\fuse \Gr{\SingIm{p_+,s;n}}$, with $s \neq p_-$.  First, we recall that $\SingIm{p_+,s;n} = \SingIm{p_+,s;n}^-$, so the resolution \eqref{eq:ResImm} (or rather the corresponding character formula \eqref{ch:Im-}) allows us to write\footnote{Note that \eqref{ch:Im-} gives two character formulae, one for $n \geqslant 0$ and the other for $n \leqslant -1$.  We assume here that $n \geqslant 0$ for clarity.  One can easily check that we get the same answer for the Verlinde product when we use the formula for $n \leqslant -1$ instead.}
\begin{align}
\Gr{\SingIm{2,1;0}} \fuse \Gr{\SingIm{p_+,s;n}} &= \sum_{k \geqslant 0} \Gr{\SingIm{2,1;0}} \fuse \brac{\Gr{\FF{p_+,p_- - s;n+2k+1}} - \Gr{\FF{p_+,s;n+2k+2}}} \nonumber \\
&= \sum_{k \geqslant 0} \left( \Gr{\FF{p_+ + 1,p_- - s;n+2k+1}} + \Gr{\FF{p_+ - 1,p_- - s;n+2k+1}} \right. \nonumber \\
&\mspace{130mu} \left. - \Gr{\FF{p_+ + 1,s;n+2k+2}} - \Gr{\FF{p_+ - 1,s;n+2k+2}} \right) \nonumber \\
\intertext{(using \eqnref{eq:ImTimesFF} and $\alpha_{r+p_+,s;n} = \alpha_{r,s;n-1}$)}
&= \sum_{k \geqslant 0} \left( \Gr{\FF{1,p_- - s;n+2k}} - \Gr{\FF{1,s;n+2k+1}} \right) + \Gr{\SingIm{p_+ - 1,s;n}^-} \nonumber \\
&= \Gr{\SingIm{1,s;n-1}^-} + \Gr{\SingIm{p_+ - 1,s;n}^-} \nonumber \\
\intertext{(using \eqnref{ch:Im-} again)}
&= \Gr{\SingIm{1,s;n-1}}+2\:\Gr{\SingIm{p_+-1,s;n}}+\Gr{\SingIm{1,s;n+1}},
\end{align}
the last equality following from \eqref{eq:MoreIdents}.  The remaining Verlinde products of \eqref{eq:GeneratorFusion} are similarly derived.

\subsection{Presentations and the Verlinde Ring}

It is clear from \eqnref{eq:GeneratorFusion} that \(\Gr{\SingIm{1,1;\pm1}}\),
\(\Gr{\SingIm{2,1;0}}\) and \(\Gr{\SingIm{1,2;0}}\) generate all of the \(\Gr{\SingIm{r,s;n}}\), for \(1\leqslant
r\leqslant p_+\), \(1\leqslant s\leqslant p_-\) and \(n\in\mathbb{Z}\), by repeatedly taking Verlinde products with one another.  Mathematically, this implies that there is 
a ring homomorphism $\phi$ from the polynomial ring \(\mathbb{Z}[X,Y,Z,Z^{-1}]\)
to the subring of the singlet Verlinde ring \(\SingVerRing{p_+,p_-}\) that is spanned by the \(\Gr{\SingIm{r,s;n}}\).  We will denote this subring by \(\SingAtypVerRing{p_+,p_-}\), referring to it as the \emph{atypical} Verlinde ring for the singlet algebra (the only typical simples involved are the $\SingIm{p_+,p_-;n} \equiv \FF{p_+,p_-;n}$).  The homomorphism $\phi$ is defined by
\begin{equation}
 \begin{gathered}
  \phi\colon\mathbb{Z}[X,Y,Z,Z^{-1}]\lra \SingAtypVerRing{p_+,p_-}; \\
  X\longmapsto \Gr{\SingIm{2,1;0}}, \qquad 
  Y\longmapsto \Gr{\SingIm{1,2;0}}, \qquad 
  Z^{\pm1}\longmapsto \Gr{\SingIm{1,1;\pm1}}.
 \end{gathered}
\end{equation}
This map is surjective by construction.

If we restrict the Verlinde products to those of \(\Gr{\SingIm{2,1;0}}\) with
\(\Gr{\SingIm{r,1;0}}\) (or \(\Gr{\SingIm{1,2;0}}\) with \(\Gr{\SingIm{1,s;0}}\)), then we observe the familiar $\SLA{sl}{2}$-structure that may be formalised in terms of Chebyshev polynomials of the second kind, at least for $r<p_+$ ($s<p_-$).  Recall that these Chebyshev polynomials are defined recursively by\footnote{Our definition is slightly non-standard and is related to the standard Chebyshev polynomials $\tfunc{\widehat{U}_n}{X}$ by $\tfunc{U_n}{X} = \tfunc{\widehat{U}_n}{X/2}$.}
\begin{align}\label{eq:ChebRecur}
  U_{-1}(X)=0,\quad U_0(X)=1;\qquad 
  U_{n+1}(X)=XU_n(X)-U_{n-1}(X)\qquad\text{(\(n\geqslant 0\))}
\end{align}
and that they satisfy the simple multiplication formulae,
\begin{align}\label{eq:ChebMult}
  U_k(X)U_\ell(X)=\sideset{}{'}\sum_{m=\abs{k-\ell}}^{k+\ell}U_m(X),
\end{align}
where the primed summation means that the label increases in steps of $2$, not $1$.
From the Verlinde products \eqref{eq:GeneratorFusion} and the recursion relations
\eqref{eq:ChebRecur}, it is not hard to see that the ring homomorphism $\phi$ acts as
\begin{align}
  \phi(U_{r-1}(X)U_{s-1}(Y)Z^n)=\Gr{\SingIm{r,s;n}}.
\end{align}
However, the kernel of \(\phi\) is non-trivial as is evidenced by the Verlinde products with $r=p_+$ and $s=p_-$.  They imply that
\begin{align}
  \phi(U_{p_+}(X)-U_{p_+-2}(X)-Z-Z^{-1}) = \phi(U_{p_-}(Y)-U_{p_--2}(Y)-Z-Z^{-1}) = 0.
\end{align}
In fact, one can show that the kernel of \(\phi\) is generated, as an ideal, by the above polynomials, hence that we have the following explicit presentation of the atypical Verlinde ring:
\begin{align}
  \SingAtypVerRing{p_+,p_-}\cong\frac{\mathbb{Z}[X,Y,Z,Z^{-1}]}
  {\langle U_{p_+}(X)-U_{p_+-2}(X)-Z-Z^{-1},U_{p_-}(Y)-U_{p_--2}(Y)-Z-Z^{-1}\rangle}\,.
\end{align}
This presentation can be used to easily compute the Verlinde product of two \(\SingIm{r,s;n}\) modules:
\begin{subequations} \label{eq:SingVerRing}
\begin{multline}\label{eq:SingVerRingRules}
  \Gr{\SingIm{r,s;n}} \fuse \Gr{\SingIm{r',s';n'}} = 
  \sideset{}{'}\sum_{j_+=\abs{r-r'}+1}^{a_+(r+r')} \ \sideset{}{'}\sum_{j_-=\abs{s-s'}+1}^{a_-(s+s')} 
  \Gr{\SingIm{j_+,j_-;n+n'}} \\
  + \sideset{}{'}\sum_{j_+=\abs{r-r'}+1}^{a_+(r+r')} \ \sideset{}{'}\sum_{j_-=b_-(s+s')}^{s+s'-1-p_-} 
  \Bigl( \Gr{\SingIm{j_+,j_-;n+n'-1}} + \Gr{\SingIm{j_+,p_--j_-;n+n'}} + \Gr{\SingIm{j_+,j_-;n+n'+1}} \Bigr) \\
  + \sideset{}{'}\sum_{j_+=b_+(r+r')}^{r+r'-1-p_+} \ \sideset{}{'}\sum_{j_-=\abs{s-s'}+1}^{a_-(s+s')} 
  \Bigl( \Gr{\SingIm{j_+,j_-;n+n'-1}} + \Gr{\SingIm{p_+-j_+,j_-;n+n'}} + \Gr{\SingIm{j_+,j_-;n+n'+1}} \Bigr) \\
  + \sideset{}{'}\sum_{j_+=b_+(r+r')}^{r+r'-1-p_+} \ \sideset{}{'}\sum_{j_-=b_-(s+s')}^{s+s'-1-p_-} 
  \Bigl( \Gr{\SingIm{j_+,j_-;n+n'-2}} + \Gr{\SingIm{j_+,p_--j_-;n+n'-1}} + \Gr{\SingIm{p_+-j_+,j_-;n+n'-1}} + 2 \: \Gr{\SingIm{j_+,j_-;n+n'}} \Bigr. \\
  \Bigl. + \Gr{\SingIm{p_+-j_+,p_--j_-;n+n'}} + \Gr{\SingIm{j_+,p_--j_-;n+n'+1}} + \Gr{\SingIm{p_+-j_+,j_-;n+n'+1}} + \Gr{\SingIm{j_+,j_-;n+n'+2}} \Bigr),
\end{multline}
where
\begin{align} \label{eq:ExplainingAB}
  a_\pm(t)=
  \begin{cases}
    t-1&\text{if \(t-1-p_\pm \leqslant 0\),}\\
    p_\pm&\text{if \(t-1-p_\pm > 0\) is even,}\\
    p_\pm-1&\text{if \(t-1-p_\pm > 0\) is odd,}
  \end{cases}
  \qquad
  b_\pm(t)=
  \begin{cases}
    1&\text{if \(t-1-p_\pm\) is odd,}\\
    2&\text{if \(t-1-p_\pm\) is even.}
  \end{cases}
\end{align}
\end{subequations}
Finally, we remark that the sums involving $b_\pm(t)$ in this result should be understood to vanish whenever $t-1-p_\pm \leqslant 0$.
This formula follows directly from the multiplication formulae \eqref{eq:ChebMult}
and the easily derived relations
\begin{equation}
\begin{aligned}
  U_{p_++k}(X)&=U_{p_+-2-k}(X)+U_k(X)(Z+Z^{-1}) & &\bmod\ker\phi & &\text{(\(0\leqslant k\leqslant p_+-2\)),} \\
  U_{p_-+k}(Y)&=U_{p_--2-k}(Y)+U_k(Y)(Z+Z^{-1}) & &\bmod\ker\phi & &\text{(\(0\leqslant k\leqslant p_--2\)).}
\end{aligned}
\end{equation}

\section{The Verlinde Ring for Triplet Models} \label{sec:TripFus}

Having determined explicit formulae for the Verlinde products of simple
singlet modules, we now consider analogous formulae for the triplet modules.
We note that there is almost nothing in the literature devoted to fusion rules
for singlet models, but that there are many sources where triplet fusion rules
have been conjectured or computed
\cite{Gaberdiel:1996np,Fuchs:2003yu,FeiLog06,Pearce:2008nn,Rasmussen:2008ez,Wood:2009ub,TsuTen12}.
Comparing these results with the triplet Verlinde products that we will deduce therefore gives very strong consistency checks of both our results and those in the literature.

\subsection{Simple Currents for the Singlet Verlinde Ring} \label{sec:SimpCurr}

As previewed in \secref{sec:Trip}, the $\Gr{\SingIm{1,1;n}}$ are simple
currents in the Verlinde ring $\SingVerRing{p_+,p_-}$, that is, they are units of
the Verlinde ring that act as permutations on the set of all simple modules.
Indeed, the Verlinde products \eqref{eq:ImTimesFF} and \eqref{eq:SimpCurrFusion} give
\begin{equation} \label{eq:SimpCurrProds}
\Gr{\SingIm{1,1;n}} \times \Gr{\FF{\mu}} = \Gr{\FF{\mu + n \alpha / 2}}, \qquad 
\Gr{\SingIm{1,1;n}} \times \Gr{\SingIm{r,s;n'}} = \Gr{\SingIm{r,s;n+n'}}.
\end{equation}
In particular, $\Gr{\SingIm{1,1;-n}}$ is the inverse, with respect to the Verlinde product, of $\Gr{\SingIm{1,1;n}}$ (because $\Gr{\SingIm{1,1;0}}$ is the identity).  We note that when $p_+ = 1$ and $p_- > 1$ ($p_- = 1$ and $p_+ > 1$), these simple currents are identified with the $\Gr{\SingIm{1,1;n}^-}$ ($\Gr{\SingIm{1,1;n}^+}$).  For $p_+ = p_- = 1$, the identification is rather with the $\Gr{\FF{1,1;n}}$.

When $p_+ = p_- = 1$, so the singlet algebra coincides with the Heisenberg
algebra, the $\FF{1,1;n}$ are well known to be simple currents in the
\emph{fusion} ring.  Indeed, extending $\SingAlg{1,1}$ by $\FF{1,1;2}$ leads to $\TripAlg{1,1} \cong \AKMA{sl}{2}_1$.  We conjecture that this generalises so that the $\SingIm{1,1;n}$ define simple currents, in a sense that we will shortly describe, with respect to the fusion product of the singlet algebra $\SingAlg{p_+,p_-}$.  As remarked in \secref{sec:Trip}, this conjecture is already suggested by the decomposition of the simple triplet algebra modules into singlet modules, at least for $n$ even.

Let us consider the case where $p_+ = 1$ and $p_- > 1$ (the case $p_- = 1$ and $p_+ > 1$ is analogous).  As we have noted above, there is then no Felder complex \eqref{eq:Felder-}, hence the list of simple $\SingAlg{1,p_-}$-modules given in \secref{sec:Sing} truncates to the typical $\FF{\lambda}$ and the atypicals $\SingIm{1,s;n} = \SingIm{1,s;n}^-$ with $1 \leqslant s \leqslant p_- - 1$ and $n \in \ZZ$ (the analogous modules with $s=p_-$ are typical:  $\SingIm{1,p_-;n}^- = \SingIm{1,p_-;n} = \FF{1,p_-;n}$).  In particular, there are no problematic modules $\VirIrr{r,s}$ to worry about and we have a bijective correspondence between the simple $\SingAlg{1,p_-}$-modules and their (linearly independent) representatives in the Verlinde ring.  We therefore claim that the Verlinde products \eqref{eq:SimpCurrProds} lift to \emph{fusion products} as follows:
\begin{equation}
\SingIm{1,1;n}^- \times \FF{\mu} = \FF{\mu + n \alpha / 2}, \qquad 
\SingIm{1,1;n}^- \times \SingIm{1,s;n'}^- = \SingIm{1,s;n+n'}^-.
\end{equation}
Note that $\SingIm{1,1;0}^-$ is the vacuum module of $\SingAlg{1,p_-}$.  These
singlet fusion products are consistent with the triplet fusion products that have 
appeared in the literature, though we will only verify this here at the level of the
Verlinde ring. 
Essentially, we claim that the Verlinde products guarantee that these fusion products are simple, hence that there are no possible ambiguities concerning their structure.  This is equivalent to fusion being exact on $\SingAlg{1,p_-}$-modules and the Verlinde ring, as defined above, coinciding with the Grothendieck ring of fusion.

If we accept these arguments supporting the $\SingIm{1,1;n}^-$ being simple currents, then it is a simple matter to determine the (simple) spectrum of the simple current extension.  We will do this for the group of simple currents corresponding to $n=2$ in order to compare with the known spectrum of the triplet algebra $\TripAlg{1,p_-}$.  As the simple currents act freely (under the fusion product) on the simple singlet modules, the simple extended algebra modules are realised by summing over the orbits of the simple current group:
\begin{equation}
\ExtFock{[\lambda]}{+} = \bigoplus_{k \in \ZZ} \FF{\lambda + k \alpha}, \quad 
\ExtFock{[\lambda]}{-} = \bigoplus_{k \in \ZZ} \FF{\lambda + (k+1/2) \alpha}; \qquad 
\TripIrr{1,s}{+} = \bigoplus_{k \in \ZZ} \SingIm{1,s;2k}^-, \quad 
\TripIrr{1,s}{-} = \bigoplus_{k \in \ZZ} \SingIm{1,s;2k+1}^-.
\end{equation}
Note that \(\ExtFock{[\lambda]}{+}=\ExtFock{[\lambda+\alpha/2]}{-}\), so we
may restrict \(\lambda\) to the real interval \(0 \leqslant \lambda < \alpha/2\).  We remark that because the simple
currents $\SingIm{1,1;n}^-$, with $n$ even, have integer conformal weights, it
is natural to restrict to the \emph{untwisted} extended algebra modules upon
which the simple current fields act with integer moding.  Referring to the
conformal weights listed in \appref{app:Structures}, we quickly find that all
of the $\TripIrr{1,s}{\pm}$ are untwisted, but that the simple
$\ExtFock{[\lambda]}{\pm}$ are only untwisted when $\lambda =
\alpha_{1,p_-}$.  Defining
\begin{equation}
\TripIrr{1,p_-}{\pm} = \ExtFock{[\alpha_{1,p_-}]}{\pm}
\end{equation}
and comparing with the decompositions of the simple triplet modules given in \secref{sec:Trip}, we conclude that the simple untwisted modules of the simple current extension of $\SingAlg{1,p_-}$ by the group generated by $\SingIm{1,1;2}^-$ may be identified with the simple $\TripAlg{1,p_-}$-modules.  We view this as extremely strong evidence for the claim that the triplet algebra is just the simple current extension of the singlet algebra by $\SingIm{1,1;2}^-$.

The general case, where $p_+, p_- > 1$, is somewhat more delicate because of the simple $\SingAlg{p_+,p_-}$-modules $\VirIrr{r,s}$ which are set to zero in the Verlinde ring.  In particular, we must allow for the possibility that a given fusion product may have composition factors of the form $\VirIrr{r,s}$ which are not visible in the Verlinde product.  This spoils any chance of a bijective correspondence between simple $\SingAlg{p_+,p_-}$-modules and their representatives in $\SingVerRing{p_+,p_-}$.  The connection between the Verlinde product and the fusion product is therefore correspondingly weaker.  The best we could hope for then is that fusion turns out to define a product on the quotient of the Grothendieck \emph{group} of $\SingAlg{p_+,p_-}$-modules by the $\VirIrr{r,s}$ and that the Verlinde ring coincides with this quotient.  We shall assume this in what follows.

Despite the expected lack of exactness and ambiguities concerning the $\VirIrr{r,s}$, we conjecture that the fusion products corresponding to the Verlinde products \eqref{eq:SimpCurrProds} are
\begin{equation} \label{eq:GenSimpCurrFus}
\begin{gathered}
\SingIm{1,1;n} \times \FF{\mu} = \FF{\mu + n \alpha / 2} \qquad \text{(\(\mu \neq \alpha_{r,s;n}\)),} \qquad \qquad 
\SingIm{1,1;n} \times \VirIrr{r,s} = 0, \\
\SingIm{1,1;n} \times \SingIm{r,s;n'}^{\bullet} = 
\begin{cases}
\bigl( \SingKer{r,s;0}^{\bullet} \bigr)^* & \text{if \(r \neq p_+\), \(s \neq p_-\) and \(n+n'=0\),} \\
\SingIm{r,s;n+n'}^{\bullet} & \text{otherwise,}
\end{cases}
\end{gathered}
\end{equation}
where ``$\bullet$'' stands for $+$, $-$ or is empty, and ``$*$'' denotes the
contragredient dual.  The appearance of the contragredient modules when $n+n'=0$ is suggested by the
$\TripAlg{2,3}$ fusion rules computed in \cite{GabFus09} and is consistent
with the $\TripAlg{p_+,p_-}$ fusion rules proposed in
\cite{Rasmussen:2008ez,Rasmussen:2009zt,Wood:2009ub}.
An interesting consequence of this is that the simple currents
$\SingIm{1,1;n}$ do not generate a group under fusion because $\bigl(
\SingKer{1,1;0}^{\bullet} \bigr)^*$ is not the vacuum module, but its
contragredient.  Instead, they generate semigroups (this possibility seems to have been first noticed in \cite{DonSim96}).

We therefore conjecture that the triplet algebra $\TripAlg{p_+,p_-}$, with $p_+, p_- > 1$, is the simple current extension of the singlet algebra $\SingAlg{p_+,p_-}$ by the semigroups generated by $\SingIm{1,1;\pm2}$.  This is easily checked to be consistent with the decompositions of the simple triplet modules given in \secref{sec:Trip} and, again, we identify the triplet modules with the untwisted modules of the simple current extension.  Evidently, this subtlety of semigroups is irrelevant at the level of the Verlinde ring, so we will assume the standard simple current machinery in what follows for all $p_+$ and $p_-$.

\subsection{Verlinde Products}

In this section, we will calculate the Verlinde products of the simple 
triplet modules under the assumption that the triplet algebra \(\TripAlg{p_+,p_-}\) 
is the simple current extension of the singlet algebra \(\SingAlg{p_+,p_-}\).  
The results are then compared with the literature, in particular with
the products proposed in \cite{FeiLog06,Rasmussen:2008ez,Rasmussen:2009zt,Wood:2009ub}. 

If this assumption is valid, then the Verlinde products of the triplet Verlinde ring
\(\TripVerRing{p_+,p_-}\) can be computed in terms of singlet Verlinde products by regarding each 
triplet module as a direct sum over an orbit of singlet modules under the action of the 
simple current (semi)group, choosing arbitrary representatives of each orbit, 
computing the Verlinde product of the representatives and, finally, 
determining the orbit of the resulting product.
It is not hard to see that this 
general procedure reduces to the following simple rules:
\begin{align}
  \Gr{\TripIrr{r,s}{+}}\extfuse\Gr{\TripIrr{r',s'}{\epsilon}}
  =\Gr{\SingIm{r,s;0}}\fuse\Gr{\TripIrr{r',s'}{\epsilon}}, \qquad 
  \Gr{\TripIrr{r,s}{-}}\extfuse\Gr{\TripIrr{r',s'}{\epsilon}}
  =\Gr{\SingIm{r,s;1}}\times\Gr{\TripIrr{r',s'}{\epsilon}}.
\end{align}
Here, we distinguish the Verlinde product ($\extfuse$) of the triplet Verlinde ring \(\TripVerRing{p_+,p_-}\) 
from that ($\fuse$) of its singlet counterpart \(\SingVerRing{p_+,p_-}\).
It is clear now that the triplet Verlinde products can be directly read off from those 
of the singlet Verlinde ring \(\SingVerRing{p_+,p_-}\). For example,
\begin{align}
  \Gr{\TripIrr{1,1}{-}}\extfuse\Gr{\TripIrr{r,s}{+}}
  &=\Gr{\SingIm{1,1;1}}\fuse \Gr{\TripIrr{r,s}{+}}
  =\Gr{\SingIm{1,1;1}}\fuse \sum_{n\in\mathbb{Z}}
  \Gr{\SingIm{r,s;2n}}\nonumber\\
  &=\sum_{n\in\mathbb{Z}}
  \Gr{\SingIm{r,s;2n+1}}=\Gr{\TripIrr{r,s}{-}}\,.
\end{align}
This shows, of course, that $\Gr{\TripIrr{1,1}{-}}$ is a simple current in the
triplet Verlinde ring --- this is the residual simple current symmetry after
extending the singlet Verlinde ring by $\SingIm{1,1;\pm2}$.\footnote{The fact
that $\TripIrr{1,1}{-}$ fuses with itself to give the \emph{contragredient}
of the vacuum module, at least for small $p_+$ and $p_-$ \cite{GabFus09,Wood:2009ub}, is our main reason for proposing the appearance of the contragredient modules in \eqref{eq:GenSimpCurrFus}.}

The general product formulae for the triplet Verlinde ring
\(\TripVerRing{p_+,p_-}\) now follow directly from the decompositions 
of \secref{sec:Trip} and the Verlinde products of the singlet Verlinde 
ring given in \eqref{eq:SingVerRingRules}.  The orbit of the singlet module $\SingIm{1,1;0}$ gives rise to the identity of $\TripVerRing{p_+,p_-}$:  $\Gr{\TripIrr{1,1}{+}}$.  As the orbits of the singlet generators $\SingIm{1,1;\pm1}$, $\SingIm{2,1;0}$ and $\SingIm{1,2;0}$ define the triplet modules $\TripIrr{1,1}{-}$, $\TripIrr{2,1}{+}$ and $\TripIrr{1,2}{+}$, respectively, \eqnref{eq:GeneratorFusion} implies their Verlinde products:
\begin{subequations}
\begin{align}
  \Gr{\TripIrr{1,1}{-}} \extfuse \Gr{\TripIrr{r,s}{\eps}} &= \Gr{\TripIrr{r,s}{-\eps}}, \\
  \Gr{\TripIrr{2,1}{+}} \extfuse \Gr{\TripIrr{r,s}{\eps}} &=
    \begin{cases}
      \Gr{\TripIrr{2,s}{\eps}} & \text{if \(r=1\),}\\
      \Gr{\TripIrr{r-1,s}{\eps}} + \Gr{\TripIrr{r+1,s}{\eps}} & \text{if \(1< r< p_+\),}\\
      2 \: \Gr{\TripIrr{p_+-1,s}{\eps}} + 2 \: \Gr{\TripIrr{1,s}{-\eps}} & \text{if \(r=p_+\),}
    \end{cases}
  \\
  \Gr{\TripIrr{1,2}{+}} \extfuse \Gr{\TripIrr{r,s}{\eps}} &=
    \begin{cases}
      \Gr{\TripIrr{r,2}{\eps}} & \text{if \(s=1\),}\\
      \Gr{\TripIrr{r,s-1}{\eps}} + \Gr{\TripIrr{r,s+1}{\eps}} & \text{if \(1< s< p_-\),}\\
      2 \: \Gr{\TripIrr{r,p_--1}{\eps}} + 2 \: \Gr{\TripIrr{r,1}{-\eps}} & \text{if \(s=p_-\).}
    \end{cases}
\end{align}
\end{subequations}
The $\Gr{\TripIrr{1,1}{-}}$, $\Gr{\TripIrr{2,1}{+}}$ and $\Gr{\TripIrr{1,2}{+}}$ therefore generate $\TripVerRing{p_+,p_-}$.  The general formula for the Verlinde product in the triplet Verlinde ring is similarly obtained from \eqref{eq:SingVerRingRules}:
\begin{multline}\label{eq:TripVerRingRules}
  \Gr{\TripIrr{r,s}{\eps}} \extfuse \Gr{\TripIrr{r',s'}{\eps'}} = 
  \sideset{}{'}\sum_{j_+=\abs{r-r'}+1}^{a_+(r+r')} \ \sideset{}{'}\sum_{j_-=\abs{s-s'}+1}^{a_-(s+s')} 
  \Gr{\TripIrr{j_+,j_-}{\eps\eps'}} \\
  + \sideset{}{'}\sum_{j_+=\abs{r-r'}+1}^{a_+(r+r')} \ \sideset{}{'}\sum_{j_-=b_-(s+s')}^{s+s'-1-p_-} 
  \Bigl( 2 \: \Gr{\TripIrr{j_+,j_-}{-\eps\eps'}} + \Gr{\TripIrr{j_+,p_--j_-}{\eps\eps'}} \Bigr)
  + \sideset{}{'}\sum_{j_+=b_+(r+r')}^{r+r'-1-p_+} \ \sideset{}{'}\sum_{j_-=\abs{s-s'}+1}^{a_-(s+s')} 
  \Bigl( 2 \: \Gr{\TripIrr{j_+,j_-}{-\eps\eps'}} + \Gr{\TripIrr{p_+-j_+,j_-}{\eps\eps'}} \Bigr) \\
  + \sideset{}{'}\sum_{j_+=b_+(r+r')}^{r+r'-1-p_+} \ \sideset{}{'}\sum_{j_-=b_-(s+s')}^{s+s'-1-p_-} 
  \Bigl( 4 \: \Gr{\TripIrr{j_+,j_-}{\eps\eps'}} + 2 \: \Gr{\TripIrr{j_+,p_--j_-}{-\eps\eps'}} + 2 \: \Gr{\TripIrr{p_+-j_+,j_-}{-\eps\eps'}} + \Gr{\TripIrr{p_+-j_+,p_--j_-}{\eps\eps'}} \Bigr),
\end{multline}
where $a_{\pm}(t)$ and $b_{\pm}(t)$ were defined in \eqref{eq:ExplainingAB}.  These formulae reproduce the Grothendieck fusion rules conjectured in \cite{FeiLog06} from a Kazhdan-Lusztig-like correspondence with a certain quantum group, and are consistent with the fusion rules proposed in \cite{Rasmussen:2008ez,Rasmussen:2009zt} from lattice considerations, and those computed for certain small values of $p_+$ and $p_-$ in \cite{Wood:2009ub} using the Nahm-Gaberdiel-Kausch algorithm, once the $\VirIrr{r,s}$ have been set to zero.

\section{Discussion and Conclusions} \label{sec:Conc}

We have seen above that the Verlinde ring of the singlet algebra
$\SingAlg{p_+,p_-}$ may be straightforwardly derived from the modular
transformation properties of the simple singlet modules and a continuous
version of the Verlinde formula.  Moreover, the
Verlinde ring of the triplet algebra $\TripAlg{p_+,p_-}$ then follows from some basic simple current technology and the result compares favourably with what is known of the triplet fusion ring.  Indeed, it appears that this rather effortless approach captures pretty much all the information about the fusion ring that can be divined from the simple characters alone.  The most difficult step was, in a sense, understanding the representation theory of the singlet algebra in the first place.

On a heuristic level, we can understand the good behaviour of the modular properties of the singlet characters, as compared with those of the triplet characters, as stemming from the uncountable nature of the spectrum of simple singlet modules.  For the singlet, the parametrisation of standard modules defines a countable set of points --- points of atypicality let us say --- at which the standard singlet modules become reducible.  For the triplet, with its finite spectrum of simple modules, one finds instead that the majority correspond to atypical points (the exception is $\TripAlg{1,1}$ of course).  Now, observe that the $\modS$-matrix elements derived for the atypical simple singlet modules have \emph{poles} at atypical parameter values.  Consequently, we see that these poles need not cause problems when integrating over a continuous spectrum as we do for the singlet (the poles form a set of measure zero after all) but that they will definitely cause problems if one tries to perform a discrete sum as one would like to do for the triplet.

It was first discovered in \cite{GabFus09} that the fusion product of the triplet
algebra $\TripAlg{p_+,p_-}$, with $p_+$ and $p_-$ greater than $1$, does not 
necessarily map exact sequences to exact sequences.  However,
this failure of exactness was always observed to involve the simple modules 
\(\VirIrr{r,s}\). As one can see from the fusion products proposed in
\cite{Rasmussen:2009zt,Wood:2009ub}, the \(\VirIrr{r,s}\) form an ideal with
respect to the fusion product in the category of \(\TripAlg{p_+,p_-}\)-modules. 
It was therefore conjectured in \cite{TsuExt13} that if one takes the
quotient of the category of \(\TripAlg{p_+,p_-}\)-modules by the ideal
generated by the \(\VirIrr{r,s}\), then the fusion product on this new
quotient category, there referred to as the Whittaker category, will be exact.  
It seems natural to expect that this quotient category is equivalent to the category 
of quantum group modules that has been studied under the name ``Kazhdan-Lusztig 
correspondence'' \cite{FeiKaz06}.

We can of course form the same kind of quotient category for the singlet
algebra \(\SingAlg{p_+,p_-}\). This quotient appears to be exactly
what the singlet Verlinde ring \(\SingVerRing{p_+,p_-}\) sees, since the
formalism we introduced above naturally sets all the \(\VirIrr{r,s}\) to
zero. We therefore conjecture that the \(\SingIm{1,1;n}\) form simple currents
of this quotient category.  Settling these conjectures would appear, to us, to
be the natural next step to tackle in understanding the $(p_+,p_-)$-models.  In 
particular, we would like to see results concerning rigidity and projectivity, 
which have received some attention for the triplet algebras, being generalised 
to the singlet algebras.  It would be interesting to know if there is some 
variant of a Kazhdan-Lusztig correspondence that applies to singlet models, 
given that these theories are arguably more fundamental in the sense that the
singlet theories are more closely related to other logarithmic theories
\cite{AdaCon05,RidSL210,CreWAl11,CreCos13} than the triplet theories.

Finally, let us remark upon some small overlap of our results with those of the recent paper \cite{CreFal13}.  There, the focus is on the $(1,p)$ singlet algebras and the relation between atypical singlet modules and the modular properties (or lack thereof) of certain variants of Jacobi theta functions that are known in number-theoretic circles as \emph{false} and \emph{partial} theta functions.  The idea is to regularise these functions and analyse modular aspects of the regularisation.  In this respect, their regularisation parameter $\eps$ plays the same role, roughly speaking, as our formal variable $z$.  They finish by computing a regularised Verlinde formula for $(1,p)$ singlet models that agrees with our results.  It would be interesting to study their regularisation procedure for general $(p_+,p_-)$ singlet algebras to see whether any of the ``bad'' features of these models, such as non-exactness of fusion, can at all be ameliorated. We suspect that the answer will be ``no'' because the geometric sum formulae being regularised in \cite{CreFal13} may instead be interpreted as identities of distributions where the test functions are linear combinations of the standard characters.

\section*{Acknowledgements}

DR's research is supported by an Australian Research Council Discovery Project DP1093910.  
SW's work is supported by the World Premier International Research Center Initiative (WPI Initiative), MEXT, Japan; the Grant-in-Aid for JSPS Fellows number 2301793; and the JSPS fellowship for foreign researchers P11793.

\appendix

\section{Structural Data} \label{app:Structures}

In this appendix, we quote the structural results concerning Feigin-Fuchs modules that are required in the text.  This material may be found in \cite{FeiRep90,IohRep11}.  We also need the corresponding results for the simple modules of the singlet algebras.  Our presentation follows \cite{TsuExt13} rather closely.

Recall that a Heisenberg weight \(\lambda\in\mathbb{R}\) and its corresponding
conformal weight \(\Delta_\lambda\) are related by \eqnref{eq:FockCW}:
\begin{align}
  \Delta_{\lambda} = \frac{1}{2} \lambda \brac{\lambda - \alpha_0} = 
  \frac{1}{2} \brac{\lambda - \frac{\alpha_0}{2}}^2 - \frac{\alpha_0^2}{8}.
\end{align}
For \(\lambda=\alpha_{r,s;n}\), we suppress the \(\alpha\) in
\(\Delta_{\alpha_{r,s;n}}\) and write
\begin{align}
  \Delta_{\alpha_{r,s;n}}\equiv\Delta_{r,s;n}=\frac{(p_-r -p_+s-2np_+p_-)^2-(p_+-p_-)^2}{4p_+p_-}.
\end{align}
We will denote the simple Virasoro module generated by a highest weight state of
central charge \(c = 1 - 3 \alpha_0^2\) and conformal weight \(\Delta\) by \(\VirIrrNM{\Delta}\).

The \emph{socle} of a Virasoro module \(M\) is its maximal semisimple submodule.  The \emph{socle series} of \(M\) is the ascending series
\begin{align}
  0=M_0\subset M_1 \subset M_2\subset\cdots
\end{align}
of submodules of \(M\) for which \(S_i(M)=M_i/M_{i-1}\) (for \(i\geqslant 1\))
is the socle of \(M_{i+1}/M_{i-1}\).
Socle series are unique if they exist.
Recalling the notation $\FF{r,s;n} \equiv \FF{\alpha_{r,s;n}}$, there are five different possibilities for the socle series factors $S_i(\FF{\lambda})$ of the Feigin-Fuchs modules \(\FF{\lambda}\):
\begin{enumerate}[leftmargin=*]
\item For \(1\leqslant r< p_+\), \(1\leqslant s<p_-\) and \(n\in\mathbb{Z}\), we have
\begin{equation}
  \begin{gathered}
    S_3(\FF{r,s;n})=\bigoplus_{k\geqslant 0}\VirIrrNM{\Delta_{p_+-r,s;|n|+2k+1}}, \\
    S_2(\FF{r,s;n})=\bigoplus_{k\geqslant a}\VirIrrNM{\Delta_{r,s;|n|+2k}}\oplus
    \bigoplus_{k\geqslant 1-a}\VirIrrNM{\Delta_{p_+-r,p_--s;|n|+2k}}, \\
    S_1(\FF{r,s;n})=\bigoplus_{k\geqslant 0}\VirIrrNM{\Delta_{r,p_--s;|n|+2k+1}},   
  \end{gathered}
\end{equation}
where \(a=0\) if \(n\geqslant 0\) and \(a=1\) if \(n<0\).
\item For \(1\leqslant s<p_-\) and \(n\in\mathbb{Z}\), we have
\begin{equation}
  \begin{gathered}
    S_2(\FF{p_+,s;n})=\bigoplus_{k\geqslant a}\VirIrrNM{\Delta_{p_+,s;|n|+2k}}, \\
    S_1(\FF{p_+,s;n})=\bigoplus_{k\geqslant 0}\VirIrrNM{\Delta_{p_+,p_--s;|n|+2k+1}},    
  \end{gathered}
\end{equation}
  where \(a=0\) if \(n\geqslant 1\) and \(a=1\) if \(n<1\).
\item For \(1\leqslant r< p_+\) and \(n\in\mathbb{Z}\), we have
\begin{equation}
  \begin{gathered}
    S_2(\FF{r,p_-;n})=\bigoplus_{k\geqslant a}\VirIrrNM{\Delta_{p_+-r,p_-;|n|+2k-1}}, \\
    S_1(\FF{r,p_-;n})=\bigoplus_{k\geqslant 0}\VirIrrNM{\Delta_{r,p_-;|n|+2k}},
  \end{gathered}
\end{equation}
  where \(a=1\) if \(n\geqslant 1\) and \(a=0\) if \(n<1\).
\item For \(n\in\mathbb{Z}\), the Feigin-Fuchs module \(\FF{p_+,p_-;n}\) is
  semisimple as a Virasoro module:
  \begin{align}
    S_1(\FF{p_+,p_-;n})=\bigoplus_{k\geqslant 0}\VirIrrNM{\Delta_{p_+,p_-;|n|+2k}}.
  \end{align}
\item Finally, for \(\lambda\in\mathbb{R}\), with \(\lambda\neq \alpha_{r,s;n}\) for any 
  \(1\leqslant r\leqslant p_+\), \(1\leqslant s\leqslant p_-\) and \(n\in\mathbb{Z}\), the
  Feigin-Fuchs module \(\FF{\lambda}\) is simple as a Virasoro module:
  \begin{align}
    S_1(\FF{\lambda})=\VirIrrNM{\Delta_\lambda}.
  \end{align}
\end{enumerate}

We will also need socle factors for the simple $\SingAlg{p_+,p_-}$-modules decomposed as Virasoro modules.  The \(\FF{\lambda}\), with \(\lambda\neq \alpha_{r,s;n}\) for any \(1\leqslant r\leqslant p_+\), \(1\leqslant s\leqslant p_-\) and \(n\in\mathbb{Z}\), are simple singlet modules, as are the \(\FF{p_+,p_-;n}\).  Their socle factors were given above.  As the result for the simple singlet modules $\VirIrr{r,s}$ is obvious, it only remains to list the factors for the $\SingIm{r,s;n}^{\bullet}$:
\begin{enumerate}[leftmargin=*]
\item For \(1\leqslant r< p_+\), \(1\leqslant s<p_-\) and \(n\in\mathbb{Z}\),
  \begin{align}
    \SingIm{r,s;n}=S_1(\FF{r,s;n})=\bigoplus_{k\geqslant 0}\VirIrrNM{\Delta_{r,p_--s;|n|+2k+1}}.
  \end{align}
  The minimal conformal weight of the states of \(\SingIm{r,s;n}\) is therefore \(\Delta_{r,p_--s;|n|+1}\).
\item For \(1\leqslant s<p_-\) and \(n\in\mathbb{Z}\),
  \begin{align}
    \SingIm{p_+,s;n}^-=S_1(\FF{p_+,s;n})=\bigoplus_{k\geqslant 0}\VirIrrNM{\Delta_{p_+,p_--s;|n|+2k+1}}.
  \end{align}
  The minimal conformal weight of the states of \(\SingIm{p_+,s;n}^-\) is therefore \(\Delta_{p_+,p_--s;|n|+1}\).
\item For \(1\leqslant r< p_+\) and \(n\in\mathbb{Z}\),
  \begin{align}
    \SingIm{r,p_-;n}^+=S_1(\FF{r,p_-;n})=\bigoplus_{k\geqslant 0}\VirIrrNM{\Delta_{r,p_-;|n|+2k}}.
  \end{align}
  The minimal conformal weight of the states of \(\SingIm{r,p_-;n}^+\) is therefore \(\Delta_{r,p_-;|n|}\).
\end{enumerate}
Of course, the minimal conformal weight of the states of \(\FF{p_+,p_-;n}\) is \(\Delta_{p_+,p_-;|n|}\) and that for \(\FF{\lambda}\) is \(\Delta_{\lambda}\).

\flushleft

\begin{thebibliography}{10}

\bibitem{KauExt91}
H~Kausch.
\newblock {Extended Conformal Algebras Generated by a Multiplet of Primary
  Fields}.
\newblock {\em Phys. Lett.}, B259:448--455, 1991.

\bibitem{GurLog93}
V~Gurarie.
\newblock {Logarithmic Operators in Conformal Field Theory}.
\newblock {\em Nucl. Phys.}, B410:535--549, 1993.
\newblock \textsf{arXiv:\mbox{hep-th}/9303160}.

\bibitem{KauCur95}
H~Kausch.
\newblock {Curiosities at $c=-2$}.
\newblock \textsf{arXiv:\mbox{hep-th}/9510149}.

\bibitem{FloMod96}
M~Flohr.
\newblock {On Modular Invariant Partition Functions of Conformal Field Theories
  with Logarithmic Operators}.
\newblock {\em Int. J. Mod. Phys.}, A11:4147--4172, 1996.
\newblock \textsf{arXiv:\mbox{hep-th}/9509166}.

\bibitem{Gaberdiel:1996np}
M~Gaberdiel and H~Kausch.
\newblock {A Rational Logarithmic Conformal Field Theory}.
\newblock {\em Phys. Lett.}, B386:131--137, 1996.
\newblock \textsf{arXiv:\mbox{hep-th}/9606050}.

\bibitem{Gaberdiel:1998ps}
M~Gaberdiel and H~Kausch.
\newblock {A Local Logarithmic Conformal Field Theory}.
\newblock {\em Nucl. Phys.}, B538:631--658, 1999.
\newblock \textsf{arXiv:\mbox{hep-th}/9807091}.

\bibitem{Fuchs:2003yu}
J~Fuchs, S~Hwang, A~Semikhatov, and I~Yu Tipunin.
\newblock {Nonsemisimple Fusion Algebras and the Verlinde Formula}.
\newblock {\em Comm. Math. Phys.}, 247:713--742, 2004.
\newblock \textsf{arXiv:\mbox{hep-th}/0306274}.

\bibitem{FeiLog06}
B~Feigin, A~Gainutdinov, A~Semikhatov, and I~Yu Tipunin.
\newblock {Logarithmic Extensions of Minimal Models: Characters and Modular
  Transformations}.
\newblock {\em Nucl. Phys.}, B757:303--343, 2006.
\newblock \textsf{arXiv:\mbox{hep-th}/0606196}.

\bibitem{NagTri11}
K~Nagatomo and A~Tsuchiya.
\newblock {The Triplet Vertex Operator Algebra $W \left( p \right)$ and the
  Restricted Quantum Group $\overline{U}_q \left( sl_2 \right)$ at $q =
  e^{\frac{\pi i}{p}}$}.
\newblock {\em Adv. Stud. Pure Math.}, 61:1--49, 2011.
\newblock \textsf{arXiv:0902.4607 [math.QA]}.

\bibitem{RasWEx09}
J~Rasmussen.
\newblock {W-Extended Logarithmic Minimal Models}.
\newblock {\em Nucl. Phys.}, B807:495--533, 2009.
\newblock \textsf{arXiv:0805.2991 [\mbox{hep-th}]}.

\bibitem{GabFus09}
M~Gaberdiel, I~Runkel, and S~Wood.
\newblock {Fusion Rules and Boundary Conditions in the $c=0$ Triplet Model}.
\newblock {\em J. Phys.}, A42:325403, 2009.
\newblock \textsf{arXiv:0905.0916 [\mbox{hep-th}]}.

\bibitem{Wood:2009ub}
S~Wood.
\newblock {Fusion Rules of the $W \left( p,q \right)$ Triplet Models}.
\newblock {\em J. Phys.}, A43:045212, 2010.
\newblock \textsf{arXiv:0907.4421 [\mbox{hep-th}]}.

\bibitem{AdaExp12}
D~Adamovi\'{c} and A~Milas.
\newblock {An Explicit Realization of Logarithmic Modules for the Vertex
  Operator Algebra $W_{p,p'}$}.
\newblock {\em J. Math. Phys.}, 53:073511, 2012.
\newblock \textsf{arXiv:1202.6667 [math.QA]}.

\bibitem{Gaberdiel93}
M~Gaberdiel.
\newblock {Fusion in Conformal Field Theory as the Tensor Product of the
  Symmetry Algebra}.
\newblock {\em Int. J. Mod. Phys.}, A9:4619--4636, 1994.
\newblock \textsf{arXiv:\mbox{hep-th}/9307183}.

\bibitem{NahQua94}
W~Nahm.
\newblock {Quasirational Fusion Products}.
\newblock {\em Int. J. Mod. Phys.}, B8:3693--3702, 1994.
\newblock \textsf{arXiv:\mbox{hep-th}/9402039}.

\bibitem{Gaberdiel:1996kx}
M~Gaberdiel and H~Kausch.
\newblock {Indecomposable Fusion Products}.
\newblock {\em Nucl. Phys.}, B477:293--318, 1996.
\newblock \textsf{arXiv:\mbox{hep-th}/9604026}.

\bibitem{VerFus88}
E~Verlinde.
\newblock {Fusion Rules and Modular Transformations in 2D Conformal Field
  Theory}.
\newblock {\em Nucl. Phys.}, B300:360--376, 1988.

\bibitem{MooPol88}
G~Moore and N~Seiberg.
\newblock {Polynomial Equations for Rational Conformal Field Theories}.
\newblock {\em Phys. Lett.}, B212:451--460, 1988.

\bibitem{MiyMod04}
M~Miyamoto.
\newblock {Modular Invariance of Vertex Operator Algebras Satisfying
  $C_2$-Cofiniteness}.
\newblock {\em Duke Math. J.}, 122:51--91, 2004.
\newblock \textsf{arXiv:math/0209101 [math.QA]}.

\bibitem{FloLog06}
M~Flohr and M~Gaberdiel.
\newblock {Logarithmic Torus Amplitudes}.
\newblock {\em J. Phys.}, A39:1955--1968, 2006.
\newblock \textsf{arXiv:\mbox{hep-th}/0509075}.

\bibitem{FloVer07}
M~Flohr and H~Knuth.
\newblock {On Verlinde-Like Formulas in $c \left( p , 1 \right)$ Logarithmic
  Conformal Field Theories}.
\newblock \textsf{arXiv:0705.0545 [\mbox{math-ph}]}.

\bibitem{Gaberdiel:2007jv}
M~Gaberdiel and I~Runkel.
\newblock {From Boundary to Bulk in Logarithmic CFT}.
\newblock {\em J. Phys.}, A41:075402, 2008.
\newblock \textsf{arXiv:0707.0388 [\mbox{hep-th}]}.

\bibitem{GaiRad09}
A~Gainutdinov and I~Yu Tipunin.
\newblock {Radford, Drinfeld and Cardy Boundary States in $(1,p)$ Logarithmic
  Conformal Field Models}.
\newblock {\em J. Phys.}, A42:315207, 2009.
\newblock \textsf{arXiv:0711.3430 [\mbox{hep-th}]}.

\bibitem{PeaGro10}
P~Pearce, J~Rasmussen, and P~Ruelle.
\newblock {Grothendieck Ring and Verlinde Formula for the W-Extended
  Logarithmic Minimal Model $WLM \left( 1 , p \right)$}.
\newblock {\em J. Phys.}, A43:045211, 2010.
\newblock \textsf{arXiv:0907.0134 [\mbox{hep-th}]}.

\bibitem{RasFus10}
J~Rasmussen.
\newblock {Fusion Matrices, Generalized Verlinde Formulas, and Partition
  Functions in $WLM \left( 1,p \right)$}.
\newblock {\em J.Phys.}, A43:105201, 2010.
\newblock \textsf{arXiv:0908.2014 [\mbox{hep-th}]}.

\bibitem{EhoHow93}
W~Eholzer, A~Honecker, and R~H\"{u}bel.
\newblock {How Complete is the Classification of W Symmetries?}
\newblock {\em Phys. Lett.}, B308:42--50, 1993.
\newblock \textsf{arXiv:\mbox{hep-th}/9302124}.

\bibitem{AdaTri08}
D~Adamovi\'{c} and A~Milas.
\newblock {On the Triplet Vertex Algebra $\mathcal{W} \left(p\right)$}.
\newblock {\em Adv. Math.}, 217:2664--2699, 2008.
\newblock \textsf{arXiv:0707.1857 [math.QA]}.

\bibitem{AdaMil10}
D~Adamovi\'{c} and A~Milas.
\newblock {On $W$-Algebras Associated to $(2, p)$ Minimal Models and Their
  Representations}.
\newblock {\em Int. Math. Res. Not.}, 2010:3896--3934, 2010.
\newblock \textsf{arXiv:0908.4053 [math.QA]}.

\bibitem{TsuExt13}
A~Tsuchiya and S~Wood.
\newblock {On the Extended $W$-Algebra of Type $sl_2$ at Positive Rational
  Level}.
\newblock \textsf{arXiv:1302.6435 [\mbox{hep-th}]}.

\bibitem{CreLog13}
T~Creutzig and D~Ridout.
\newblock {Logarithmic Conformal Field Theory: Beyond an Introduction}.
\newblock \textsf{arXiv:1303.0847 [\mbox{hep-th}]}.

\bibitem{RozQua92}
L~Rozansky and H~Saleur.
\newblock {Quantum Field Theory for the Multivariable Alexander-Conway
  Polynomial}.
\newblock {\em Nucl. Phys.}, B376:461--509, 1992.

\bibitem{QueFre07}
T~Quella and V~Schomerus.
\newblock {Free Fermion Resolution of Supergroup WZNW Models}.
\newblock {\em JHEP}, 0709:085, 2007.
\newblock \textsf{arXiv:0706.0744 [\mbox{hep-th}]}.

\bibitem{CreRel11}
T~Creutzig and D~Ridout.
\newblock {Relating the Archetypes of Logarithmic Conformal Field Theory}.
\newblock {\em Nucl. Phys.}, B872:348--391, 2013.
\newblock \textsf{arXiv:1107.2135 [\mbox{hep-th}]}.

\bibitem{CreMod12}
T~Creutzig and D~Ridout.
\newblock {Modular Data and Verlinde Formulae for Fractional Level WZW Models
  I}.
\newblock {\em Nucl. Phys.}, B865:83--114, 2012.
\newblock \textsf{arXiv:1205.6513 [\mbox{hep-th}]}.

\bibitem{RidSL208}
D~Ridout.
\newblock {$\widehat{\mathfrak{sl}} \left( 2 \right)_{-1/2}$: A Case Study}.
\newblock {\em Nucl. Phys.}, B814:485--521, 2009.
\newblock \textsf{arXiv:0810.3532 [\mbox{hep-th}]}.

\bibitem{CreMod13}
T~Creutzig and D~Ridout.
\newblock {Modular Data and Verlinde Formulae for Fractional Level WZW Models
  II}.
\newblock {\em Nucl. Phys.}, B875:423--458, 2013.
\newblock \textsf{arXiv:1306.4388 [\mbox{hep-th}]}.

\bibitem{CreFal13}
T~Creutzig and A~Milas.
\newblock {False Theta Functions and the Verlinde Formula}.
\newblock \textsf{arXiv:1309.6037 [math.QA]}.

\bibitem{BabTak12}
A~Babichenko and D~Ridout.
\newblock {Takiff Superalgebras and Conformal Field Theory}.
\newblock {\em J. Phys.}, A46:125204, 2013.
\newblock \textsf{arXiv:1210.7094 [\mbox{math-ph}]}.

\bibitem{FeiRep90}
B~Feigin and D~Fuchs.
\newblock {Representations of the Virasoro Algebra}.
\newblock {\em Adv. Stud. Contemp. Math.}, 7:465--554, 1990.

\bibitem{IohRep11}
K~Iohara and Y~Koga.
\newblock {\em {Representation Theory of the Virasoro Algebra}}.
\newblock Springer Monographs in Mathematics. Springer-Verlag, London, 2011.

\bibitem{FelBRST89}
G~Felder.
\newblock {BRST Approach to Minimal Models}.
\newblock {\em Nucl. Phys.}, B317:215--236, 1989.

\bibitem{TsuKan:1986}
A~Tsuchiya and Y~Kanie.
\newblock {Fock Space Representations of the Virasoro Algebra -- Intertwining
  Operators}.
\newblock {\em Publ. RIMS, Kyoto Univ.}, 22:259--327, 1986.

\bibitem{AdaCla03}
D~Adamovi\'{c}.
\newblock {Classification of Irreducible Modules of Certain Subalgebras of Free
  Boson Vertex Algebra}.
\newblock {\em J. Alg.}, 270:115--132, 2003.
\newblock \textsf{arXiv:math/0207155 [math.QA]}.

\bibitem{AdaMil09}
D~Adamovi\'{c} and A~Milas.
\newblock {Logarithmic intertwining operators and W(2,2p-1)-algebras}.
\newblock {\em J. Math. Phys.}, 48:073503, 2007.
\newblock \textsf{arXiv:math/0702081 [math.QA]}.

\bibitem{TsuTen12}
A~Tsuchiya and S~Wood.
\newblock {The Tensor Structure on the Representation Category of the
  $\mathcal{W}_p$ Triplet Algebra}.
\newblock \textsf{arXiv:1201.0419 [\mbox{hep-th}]}.

\bibitem{Pearce:2008nn}
P~Pearce, J~Rasmussen, and P~Ruelle.
\newblock {Integrable Boundary Conditions and $W$-Extended Fusion in the
  Logarithmic Minimal Models $LM \left( 1,p \right)$}.
\newblock {\em J. Phys.}, A41:295201, 2008.
\newblock \textsf{arXiv:0803.0785 [\mbox{hep-th}]}.

\bibitem{Rasmussen:2008ez}
J~Rasmussen.
\newblock {Polynomial Fusion Rings of W-Extended Logarithmic Minimal Models}.
\newblock {\em J. Math. Phys.}, 50:043512, 2009.
\newblock \textsf{arXiv:0812.1070 [\mbox{hep-th}]}.

\bibitem{Rasmussen:2009zt}
J~Rasmussen.
\newblock {Fusion of Irreducible Modules in $WLM \left( p,p' \right)$}.
\newblock {\em J. Phys.}, A43:045210, 2010.
\newblock \textsf{arXiv:0906.5414 [\mbox{hep-th}]}.

\bibitem{DonSim96}
C~Dong, H~Li, and G~Mason.
\newblock {Simple Currents and Extensions of Vertex Operator Algebras}.
\newblock {\em Comm. Math. Phys.}, 180:671--707, 1996.
\newblock \textsf{arXiv:\mbox{q-alg}/9504008}.

\bibitem{FeiKaz06}
B~Feigin, A~Gainutdinov, A~Semikhatov, and I~Yu Tipunin.
\newblock {Kazhdan-Lusztig Correspondence for the Representation Category of
  the Triplet $W$-Algebra in Logarithmic CFT}.
\newblock {\em Theo. Math. Phys.}, 148:1210--1235, 2006.
\newblock \textsf{arXiv:math/0512621}.

\bibitem{AdaCon05}
D~Adamovi\'{c}.
\newblock {A Construction of Admissible $A_1^{\left(1\right)}$-Modules of level
  $-\frac{4}{3}$}.
\newblock {\em J. Pure Appl. Alg.}, 196:119--134, 2005.
\newblock \textsf{arXiv:math/0401023 [math.QA]}.

\bibitem{RidSL210}
D~Ridout.
\newblock {$\widehat{\mathfrak{sl}} \left( 2 \right)_{-1/2}$ and the Triplet
  Model}.
\newblock {\em Nucl. Phys.}, B835:314--342, 2010.
\newblock \textsf{arXiv:1001.3960 [\mbox{hep-th}]}.

\bibitem{CreWAl11}
T~Creutzig and D~Ridout.
\newblock {W-Algebras Extending $\widehat{\mathfrak{gl}} \left( 1 \middle\vert
  1 \right)$}.
\newblock In V~Dobrev, editor, {\em Proceedings of the 9-th International
  Workshop "Lie Theory and Its Applications in Physics"}, volume~36 of {\em
  Springer Proceedings in Mathematics and Statistics}, pages 349--368, Varna,
  2011.
\newblock \textsf{arXiv:1111.5049 [\mbox{hep-th}]}.

\bibitem{CreCos13}
T~Creutzig, D~Ridout, and S~Wood.
\newblock {Coset Constructions of Logarithmic $\left( 1,p \right)$-Models}.
\newblock \textsf{arXiv:1305.2665 [math.QA]}.

\end{thebibliography}

\end{document}